\documentclass[lettersize,journal]{IEEEtran}

\usepackage{graphicx}
\usepackage[caption=false, font=footnotesize]{subfig}
\usepackage[table]{xcolor}
\usepackage{tikz}
\usetikzlibrary{arrows.meta,positioning,shapes.geometric}
\usetikzlibrary{positioning,arrows.meta}
\usepackage{amsmath,amssymb}
\usepackage{tikz}
\usetikzlibrary{positioning,arrows.meta}
\usepackage{cite}
\usepackage{booktabs}
\usepackage{tabularx}
\usepackage{array}
\usepackage{amssymb}
\usepackage{xcolor}
\usepackage{tikz}
\usepackage{float}
\usepackage{forest}
\usepackage{xcolor}
\usepackage{forest}
\usetikzlibrary{trees}
\usepackage[hidelinks]{hyperref}
\usepackage{tabularx}
\usepackage{array}
\renewcommand{\arraystretch}{1.5}
\usepackage{algorithm}
\usepackage{algorithmicx}
\usepackage{algpseudocode}
\usepackage{multirow}
\usepackage{booktabs}

\usepackage[acronym]{glossaries}
\makeglossaries
\newacronym{mcrt}{MCRT}{Monte Carlo Radiative Transfer}
\newacronym{rte}{RTE}{Radiative Transfer Equation}
\newacronym{mc}{MC}{Monte Carlo}

\newacronym{uowc}{UOWC}{Underwater Optical Wireless Communication}
\newacronym{uow}{UOW}{Underwater Optical Wireless}
\newacronym{uowsc}{UOWSC}{Underwater Optical Wireless Semantic Communication}
\newacronym{mse}{MSE}{Mean Squared Error}
\newacronym{auvs}{AUVs}{Autonomous Underwater Vehicles}
\newacronym{lms}{LMS}{Least Mean Squares}
\newacronym{ook}{OOK}{On-Off Keying}
\newacronym{auv}{AUV}{Autonomous Underwater Vehicle}
\newacronym{lfn}{LFN}{low frequency noise}
\newacronym{imdd}{IM/DD}{intensity modulation with direct detection}
\newacronym{rq}{RQ}{Residual Quantization}
\newacronym{ekf}{EKF}{Extended Kalman Filter}
\newacronym{gps}{GPS}{Global Positioning System}
\newacronym{hdfec}{HD-FEC}{hard-decision forward error correction}
\newacronym{apsk}{APSK}{Amplitude and Phase Shift Keying}
\newacronym{ls}{LS}{least squares}
\newacronym{lmmse}{LMMSE}{linear minimum mean-square error}
\newacronym{bler}{BLER}{block error rate}
\newacronym{lbc}{LBC}{linear block code}
\newacronym{js}{JS}{Jensen--Shannon}
\newacronym{led}{LED}{light-emitting diode}
\newacronym{spl}{SPL}{symbol and power loading}
\newacronym{asv}{ASV}{Autonomous Surface Vehicle}
\newacronym{usv}{USV}{Unmanned Surface Vehicle}
\newacronym{rov}{ROV}{Remotely Operated Vehicle}
\newacronym{iout}{IoUT}{Internet of Underwater Things}
\newacronym{wgg}{WGG}{Weibull-Generalized Gamma}
\newacronym{rf}{RF}{Radio Frequency}
\newacronym{los}{LOS}{Line-of-Sight}
\newacronym{nlos}{NLOS}{Non-Line-of-Sight}
\newacronym{fov}{FOV}{Field of View}
\newacronym{psf}{PSF}{Point Spread Function}
\newacronym{gpu}{GPU}{Graphics Processing Unit}
\newacronym{cpu}{CPU}{Central Processing Unit}
\newacronym{dac}{DAC}{Digital-to-Analog Converter}
\newacronym{adc}{ADC}{Analog-to-Digital Converter}
\newacronym{spad}{SPAD}{Single-Photon Avalanche Diode}
\newacronym{apd}{APD}{Avalanche Photodiode}
\newacronym{pin}{PIN}{Positive-Intrinsic-Negative}
\newacronym{snr}{SNR}{Signal-to-Noise Ratio}
\newacronym{ber}{BER}{Bit Error Rate}
\newacronym{nmse}{NMSE}{Normalized Mean Squared Error}
\newacronym{psnr}{PSNR}{Peak Signal-to-Noise Ratio}
\newacronym{ssim}{SSIM}{Structural Similarity Index Measure}
\newacronym{ppm}{PPM}{Pulse Position Modulation}
\newacronym{ofdm}{OFDM}{Orthogonal Frequency Division Multiplexing}
\newacronym{im}{IM}{Intensity Modulation}
\newacronym{dd}{DD}{Direct Detection}
\newacronym{isi}{ISI}{Intersymbol Interference}
\newacronym{fec}{FEC}{Forward Error Correction}
\newacronym{mcs}{MCS}{Modulation and Coding Scheme}
\newacronym{qam}{QAM}{Quadrature Amplitude Modulation}
\newacronym{pam}{PAM}{Pulse Amplitude Modulation}
\newacronym{qpsk}{QPSK}{Quadrature Phase Shift Keying}
\newacronym{w2a}{W2A}{Water-to-Air}
\newacronym{8psk}{8PSK}{8-ary Phase Shift Keying}
\newacronym{4qam}{4QAM}{4-ary Quadrature Amplitude Modulation}
\newacronym{16qam}{16QAM}{16-ary Quadrature Amplitude Modulation}
\newacronym{64qam}{64QAM}{64-ary Quadrature Amplitude Modulation}
\newacronym{sagmamba}{SAGMamba}{Spatial Aware Global Mamba}
\newacronym{pmgffn}{PMGFFN}{Physical Model-Guided Feed-Forward Network}
\newacronym{owc}{OWC}{Optical Wireless Communication}
\newacronym{roc}{ROC}{Receiver Operating Characteristic}
\newacronym{ann}{ANN}{Artificial Neural Network}
\newacronym{nn}{NN}{Neural Network}
\newacronym{dnn}{DNN}{Deep Neural Network}
\newacronym{cnn}{CNN}{Convolutional Neural Network}
\newacronym{cnns}{CNNs}{Convolutional Neural Networks}
\newacronym{rnn}{RNN}{Recurrent Neural Network}
\newacronym{lstm}{LSTM}{Long Short-Term Memory}
\newacronym{gru}{GRU}{Gated Recurrent Unit}
\newacronym{mlp}{MLP}{Multilayer Perceptron}
\newacronym{dbn}{DBN}{Deep Belief Network}
\newacronym{rbm}{RBM}{Restricted Boltzmann Machine}
\newacronym{vae}{VAE}{Variational Autoencoder}
\newacronym{aae}{AAE}{Adversarial Autoencoder}
\newacronym{gan}{GAN}{Generative Adversarial Network}
\newacronym{ae}{AE}{Autoencoder}

\newacronym{ml}{ML}{Machine Learning}
\newacronym{dl}{DL}{Deep Learning}
\newacronym{maml}{MAML}{Model-Agnostic Meta-Learning}
\newacronym{svd}{SVD}{Singular Value Decomposition}
\newacronym{knn}{KNN}{K-Nearest Neighbors}
\newacronym{svm}{SVM}{Support Vector Machine}
\newacronym{sgd}{SGD}{Stochastic Gradient Descent}
\newacronym{mae}{MAE}{Mean Absolute Error}
\newacronym{rl}{RL}{Reinforcement Learning}
\newacronym{drl}{DRL}{Deep Reinforcement Learning}
\newacronym{dqn}{DQN}{Deep Q-Network}
\newacronym{ddpg}{DDPG}{Deep Deterministic Policy Gradient}
\newacronym{semitnn}{semiTNN}{Semi-supervised Twin Neural Network}
\newacronym{sarsa}{SARSA}{State-Action-Reward-State-Action}
\newacronym{mdp}{MDP}{Markov Decision Process}

\newacronym{uwpt}{UWPT}{Underwater Wireless Power Transfer}
\newacronym{slipt}{SLIPT}{Simultaneous Lightwave Information and Power Transfer}
\newacronym{ciecl}{CIECL}{Cooperative Information-Energy Capacity Learning}

\newacronym{dom}{DOM}{Discrete Ordinates Method}
\newacronym{pn}{PN}{Spherical Harmonics Method}
\newacronym{fem}{FEM}{Finite Element Method}
\newacronym{fdm}{FDM}{Finite Difference Method}
\newacronym{bem}{BEM}{Boundary Element Method}
\newacronym{disort}{DISORT}{Discrete Ordinates Radiative Transfer Solver}
\newacronym{msvs}{MSVs}{Marine Surface Vehicles}
\newacronym{jscc}{JSCC}{Joint Source-Channel Coding}
\newacronym{csi}{CSI}{Channel State Information}
\newacronym{elbo}{ELBO}{Evidence Lower Bound}
\newacronym{vlc}{VLC}{Visible Light Communication}
\newacronym{mimo}{MIMO}{Multiple-Input Multiple-Output}
\newacronym{marl}{MARL}{Multi-Agent Reinforcement Learning}
\newacronym{llr}{LLR}{Log-Likelihood Ratio}
\newacronym{oam}{OAM}{Orbital Angular Momentum}
\newacronym{oamsk}{OAM-SK}{Orbital Angular Momentum Shift Keying}
\newacronym{cgan}{CGAN}{Conditional Generative Adversarial Network}
\newacronym{dcgan}{DCGAN}{Deep Convolutional Generative Adversarial Network}
\newacronym{dccgan}{DCC-GAN}{Deep Convolutional Conditional Generative Adversarial Network}
\newacronym{rmse}{RMSE}{Root Mean Squared Error}
\newacronym{uvlc}{UVLC}{Underwater Visible Light Communication}
\newacronym{srcnn}{SRCNN}{Super-Resolution Convolutional Neural Network}
\newacronym{dncnn}{DnCNN}{Denoising Convolutional Neural Network}
\newacronym{ssbi}{SSBI}{Signal-to-Signal Beat Interference}
\newacronym{ppdnn}{PPDNN}{Partially Pruned Deep Neural Network}
\newacronym{eann}{EANN}{Energy-Adaptive Neural Network}
\newacronym{dqlaecm}{DQL-AECM}{Deep Q-Learning-based Adaptive Error Correction and Modulation}
\newacronym{uosn}{UOSN}{Underwater Optical Sensor Network}
\newacronym{wos}{WOS}{Wave Optics Simulation}
\newacronym{egg}{EGG}{Exponential-Generalized Gamma}
\newacronym{dcoofdm}{DCO-OFDM}{DC-biased Optical Orthogonal Frequency Division Multiplexing}
\newacronym{tcn}{TCN}{Temporal Convolutional Network}
\newacronym{bgru}{BGRU}{Bidirectional Gated Recurrent Unit}
\newacronym{deepesn}{DeepESN}{Deep Echo State Network}
\newacronym{apt}{APT}{Acquisition, Pointing, and Tracking}
\newacronym{sac}{SAC}{Soft Actor-Critic}
\newacronym{iop}{IOP}{Inherent Optical Properties}
\newacronym{fso}{FSO}{Free Space Optical}
\newacronym{jpeg}{JPEG}{Joint Photographic Experts Group}
\newacronym{ldpc}{LDPC}{Low-Density Parity-Check}
\newacronym{bcr}{BCR}{Bandwidth Compression Ratio}
\newacronym{bd}{BD}{Bjøntegaard Delta}

\graphicspath{{figures/}}

\begin{document}
\title{Machine Learning for Underwater Optical Wireless 
Communication Systems: A Comprehensive Survey}

\author{Shaymaa~Mahmoud,~\IEEEmembership{Student~Member,~IEEE,}
        Ardimas~Purwita,~\IEEEmembership{Member,~IEEE,}
        and~Mohamed-Slim~Alouini,~\IEEEmembership{Fellow,~IEEE}}

\maketitle
\begin{abstract}
\gls{uowc} has emerged as a promising technology for 
high-speed underwater data transmission, offering 
significantly higher bandwidth and lower latency compared to acoustic and \gls{rf} technologies. However, the underwater medium introduces severe impairments that degrade link performance and limit communication range. The growing complexity of these challenges has increased research interest in \gls{ml} and \gls{dl} approaches, which offer powerful tools for channel modeling, signal processing, and system adaptation in ways that conventional analytical methods struggle to achieve. This survey provides a comprehensive review of \gls{ml} methods applied across the full \gls{uowc} system pipeline, covering channel modeling, transmitter design, receiver detection and equalization, link alignment, and emerging applications, including wireless power transfer, semantic communication, object detection, optical sensing, and localization. Finally, we discuss open challenges and future research directions to motivate further work in this rapidly growing field.
\end{abstract}
\begin{table*}[!t]
\caption{Principal Abbreviations Used in This Survey}
\label{tab:abbreviations}
\centering
\footnotesize
\renewcommand{\arraystretch}{1.12}
% Alphabetical by abbreviation, reading down the left pair then the right.
\begin{tabularx}{\textwidth}{@{}>{\bfseries}l>{\raggedright\arraybackslash}X>{\bfseries}l>{\raggedright\arraybackslash}X@{}}
\toprule
\textbf{Abbreviation} & \textbf{Definition} & \textbf{Abbreviation} & \textbf{Definition} \\
\midrule
\glsentryshort{ae}    & \glsentrylong{ae}    & \glsentryshort{mlp}   & \glsentrylong{mlp} \\
\glsentryshort{ann}   & \glsentrylong{ann}   & \glsentryshort{nn}    & \glsentrylong{nn} \\
\glsentryshort{apd}   & \glsentrylong{apd}   & \glsentryshort{oam}   & \glsentrylong{oam} \\
\glsentryshort{apt}   & \glsentrylong{apt}   & \glsentryshort{ofdm}  & \glsentrylong{ofdm} \\
\glsentryshort{auv}   & \glsentrylong{auv}   & \glsentryshort{ook}   & \glsentrylong{ook} \\
\glsentryshort{ber}   & \glsentrylong{ber}   & \glsentryshort{owc}   & \glsentrylong{owc} \\
\glsentryshort{cnn}   & \glsentrylong{cnn}   & \glsentryshort{pam}   & \glsentrylong{pam} \\
\glsentryshort{csi}   & \glsentrylong{csi}   & \glsentryshort{qam}   & \glsentrylong{qam} \\
\glsentryshort{ddpg}  & \glsentrylong{ddpg}  & \glsentryshort{rl}    & \glsentrylong{rl} \\
\glsentryshort{dl}    & \glsentrylong{dl}    & \glsentryshort{rnn}   & \glsentrylong{rnn} \\
\glsentryshort{dnn}   & \glsentrylong{dnn}   & \glsentryshort{rov}   & \glsentrylong{rov} \\
\glsentryshort{dqn}   & \glsentrylong{dqn}   & \glsentryshort{rte}   & \glsentrylong{rte} \\
\glsentryshort{drl}   & \glsentrylong{drl}   & \glsentryshort{sac}   & \glsentrylong{sac} \\
\glsentryshort{fso}   & \glsentrylong{fso}   & \glsentryshort{slipt} & \glsentrylong{slipt} \\
\glsentryshort{gan}   & \glsentrylong{gan}   & \glsentryshort{snr}   & \glsentrylong{snr} \\
\glsentryshort{imdd}  & \glsentrylong{imdd}  & \glsentryshort{spad}  & \glsentrylong{spad} \\
\glsentryshort{led}   & \glsentrylong{led}   & \glsentryshort{uowc}  & \glsentrylong{uowc} \\
\glsentryshort{lstm}  & \glsentrylong{lstm}  & \glsentryshort{uowsc} & \glsentrylong{uowsc} \\
\glsentryshort{maml}  & \glsentrylong{maml}  & \glsentryshort{usv}   & \glsentrylong{usv} \\
\glsentryshort{mc}    & \glsentrylong{mc}    & \glsentryshort{uvlc}  & \glsentrylong{uvlc} \\
\glsentryshort{mcrt}  & \glsentrylong{mcrt}  & \glsentryshort{uwpt}  & \glsentrylong{uwpt} \\
\glsentryshort{mimo}  & \glsentrylong{mimo}  & \glsentryshort{vae}   & \glsentrylong{vae} \\
\glsentryshort{ml}    & \glsentrylong{ml}    & \glsentryshort{vlc}   & \glsentrylong{vlc} \\
\bottomrule
\end{tabularx}
\end{table*}

\glsresetall


\section{Introduction}
\label{sec:intro}
\Gls{uowc} has emerged as a promising technology for high-speed underwater data transmission, supporting critical applications including \gls{auv} navigation, ocean data collection, and underwater sensor networks. Compared to traditional acoustic and \gls{rf} systems, \gls{uowc} offers significantly lower latency and higher bandwidth~\cite{Zeng2017Survey, Kaushal2016}. Beyond laboratory studies, \gls{uowc} has reached commercial deployment in operational underwater systems: Sonardyne's BlueComm~200 achieves data rates of 2.5--10~Mbps at ranges up to 150~m and depths of 4,000~m~\cite{Sonardyne2021BlueComm200}, while Hydromea's LUMA~X modem delivers data rates of up to 10~Mbps 
at ranges up to 50~m, enabling real-time high-definition video transmission~\cite{HydromeaLUMAX}. However, the performance of \gls{uowc} systems remains fundamentally limited by the complex nature of the underwater optical channel.
\subsection{Overview of Underwater Optical Wireless Communication}
Light propagation in water is simultaneously affected by optical absorption, multiple scattering from suspended particles, turbulence-induced fading caused by temperature gradients and salinity variations, and geometric misalignment between the transmitter and receiver. These effects reduce the reliable communication range and degrade signal quality. The propagation of light in water is commonly described by the integro-differential \gls{rte}~\cite{Zaneveld2007RTE, Mobley1994}, which generally cannot be solved analytically and motivates the development of numerical solution techniques. Among these, \gls{mcrt} is regarded as one of the most accurate and widely adopted methods for \gls{uowc} channel simulation~\cite{Gabriel2013MC, mobley2022oceanic}.

The underwater optical channel exhibits highly nonlinear and nonstationary behavior. The design of reliable \gls{uowc} systems requires solving a series of interdependent problems across the entire communication pipeline: adaptive transmitter design under dynamic channel conditions, robust signal detection and equalization at the receiver, and precise beam alignment for laser links. Conventional approaches address each of these problems independently, relying on closed-form analytical expressions or fixed signal processing blocks that assume stationary channel statistics. These assumptions are frequently violated in realistic underwater deployments, where channel conditions vary substantially with water conditions and turbulence intensity~\cite{8370053}.

The rapid progress of \gls{ml} has motivated a new direction for \gls{uowc} system design. This direction is driven by several limitations of conventional approaches. Conventional analytical models such as the Beer-Lambert law struggle to accurately capture the highly nonlinear channel behavior arising from the combined 
effects of turbulence-induced fading,
absorption, and multiple scattering~\cite{s20082261,Jasman}. Variations in water conditions, turbulence intensity, and link geometry can significantly alter channel statistics, degrading the performance of pre-designed equalizers and other fixed signal processing techniques~\cite{8606206, photonics10070811}. Additionally, high-speed receivers based on \glspl{spad} introduce hardware nonlinearities that are difficult to capture using conventional models~\cite{8336942,8962099}. \gls{ml} methods address these limitations directly: \gls{dl} approaches model hardware and channel nonlinearities through their universal approximation capability, \gls{rl} can be utilized in link alignment and beam adaptation, and physical models can further be used for channel characterization. Beyond data transmission, \gls{ml} and \gls{dl} models have been used in various applications including underwater image enhancement, optical object detection, multi-sensor localization, and simultaneous lightwave information and power transfer~\cite{zhang2025nav, shin2024adaptive,tiwari2025mine}. These 
developments motivate a system-level survey of \gls{ml} methods across the full \gls{uowc} pipeline.

\begin{figure}[!t]
\centering
% Source: figures/uwoc-system-architecture.svg; regenerate the PDF with
%   rsvg-convert -f pdf -o figures/uwoc-system-architecture.pdf figures/uwoc-system-architecture.svg
\includegraphics[width=\columnwidth]{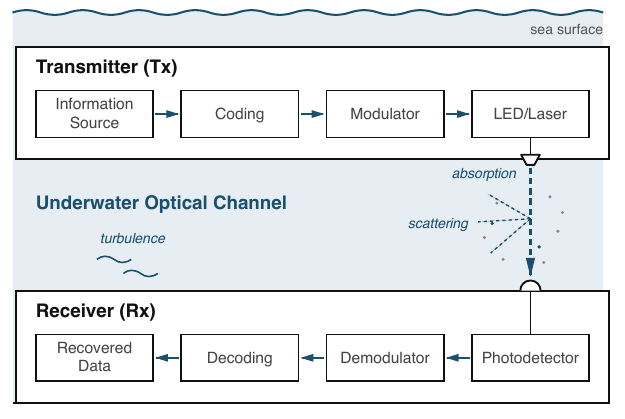}
\caption{General architecture of an underwater optical wireless communication system. Information is encoded, modulated, and converted into optical signals by an \gls{led}/laser transmitter. The optical beam
propagates through the \gls{uowc} channel, where absorption, scattering, and turbulence degrade the signal, after which photodetection, demodulation, and decoding are performed to recover the transmitted
information.}
\label{fig:uwoc_system}
\end{figure}

\subsection{ Review of Related \gls{uowc} Surveys}
\label{subsec:related}

Existing survey literature related to this work can be classified into two categories: surveys on \gls{uowc} systems and surveys on \gls{ml} for communication systems. Several surveys have reviewed the fundamentals of \gls{uowc}. Zeng~\textit{et al.}~\cite{Zeng2017Survey} and Saeed~\textit{et al.}~\cite{Saeed2019} provide comprehensive overviews of underwater optical channel characteristics, propagation impairments, modulation techniques, networking architectures, and system design considerations. More recently, a survey on \gls{uowc} channel modeling was presented in~\cite{Li2025RTEsurvey}, where the authors reviewed numerical solutions of the \gls{rte}, including deterministic methods and \gls{mcrt}-based approaches for underwater optical channel simulation. Although these surveys establish the foundations of \gls{uowc}, they primarily focus on channel modeling, propagation analysis, and communication architectures, with limited discussion of modern \gls{ml} techniques.

\Gls{ml} has been applied to physical-layer and system design in~\cite{OShea2017}. In the optical communications domain, Amirabadi \textit{et al.}~\cite{11003870} present a comprehensive survey of \gls{ml} and \gls{dl} across three optical communication technologies: optical fiber communications, optical networking, and \gls{owc}, including \gls{fso} links. The survey covers a wide range of applications including coherent transceiver design, digital signal processing, optical performance monitoring, and nonlinearity compensation. While this work provides valuable coverage of \gls{ml} in optical systems, it does not address the specific physical environment of underwater optical propagation.

Most closely related to this work, Shalol \textit{et al.}~\cite{Shalol2026} reviewed end-to-end learning for \gls{uowc} systems with a focus on \gls{ae} architectures. The survey proposed a taxonomy of
deterministic, denoising, variational, hybrid, adaptive, and physics-informed \gls{ae} variants, and benchmarked their reported \gls{ber}, \gls{snr} range, and throughput across water types and turbulence conditions. While this work provides a detailed treatment of \gls{ae}-based transceiver design, its scope is limited to a single learning framework and to the transmission chain itself. It does not cover the broader spectrum of \gls{ml} methods applied to \gls{uowc}, including \gls{rl} for link alignment and beam control and meta-learning for channel adaptation.

Gupta and Goel~\cite{10.1117/12.3108135} provide a concise review of \gls{dl} approaches for underwater optical communication, highlighting \glspl{cnn}, recurrent architectures, autoencoders, and selected learning-assisted \gls{ofdm}, \gls{oam}, and hybrid optical links. However, their review does not provide a component-wise analysis across the complete \gls{uowc} pipeline or systematically compare the datasets, channel conditions, validation settings, and performance metrics of the reviewed methods. Learning-based link alignment, semantic communication, meta learning, multi-agent reinforcement learning, and emerging sensing, localization, and power-transfer applications are also not examined in depth. Fang \textit{et al.}~\cite{photonics10070811} surveyed high-speed \gls{uowc} systems with an emphasis on advanced signal processing methods. Their review provides detailed coverage of conventional equalization and equalization using machine learning. It also discusses applications of reinforcement learning in beam adaptation, link alignment, and routing. However, its primary focus remains signal processing and equalization rather than a component-wise synthesis of learning methods across channel modeling, end-to-end and semantic transceiver design, meta-learning, datasets and validation settings, and emerging sensing, localization, and power-transfer applications.

In \gls{fso} communication, Al-Imran~\textit{et al.}~\cite{ALIMRAN20251026} provide a review of \gls{ml} and \gls{dl} techniques specifically for \gls{fso} systems, covering channel estimation, demodulation, hybrid \gls{fso}/\gls{rf} systems, and underwater \gls{fso}. Although this survey reviews several \gls{ml} methods in underwater FSO, it does not focus on the distinct physical impairments of the underwater optical channel, including multiple scattering, absorption, turbulence, and hardware constraints.  Additionally, Al Imran~\textit{et al.}~\cite{11142888} provide a review of \gls{ml} and \gls{dl} techniques in \gls{vlc}, focusing on channel estimation, noise mitigation, modulation classification, and symbol detection. While \gls{vlc} and \gls{uowc} both use optical carriers, they operate in fundamentally different environments. 
Doha and Abdelhadi~\cite{11053759} survey \gls{dl} applications in wireless communication receivers, but focus on \gls{rf} systems rather than optical underwater channels.Based on the related surveys summarized in Table~\ref{tab:related_surveys},
existing reviews do not jointly provide coverage across channel modeling, transmitter and receiver processing, link alignment, and
emerging \gls{uowc} applications within a unified \gls{ml}taxonomy.
\subsection{Contribution and Organization}
As illustrated in Table~\ref{tab:related_surveys}, this survey reviews \gls{ml} approaches for channel modeling, transmitter design, receiver design, link alignment, and other emerging \gls{uowc} applications. In addition, it analyzes the role of different \gls{ml} methods within each system component, and identifies key open challenges and future research directions.

\begin{figure*}[!t]
\centering
% Source: figures/uwoc-organization.svg; regenerate the PDF with
%   rsvg-convert -f pdf -o figures/uwoc-organization.pdf figures/uwoc-organization.svg
\includegraphics[width=\textwidth]{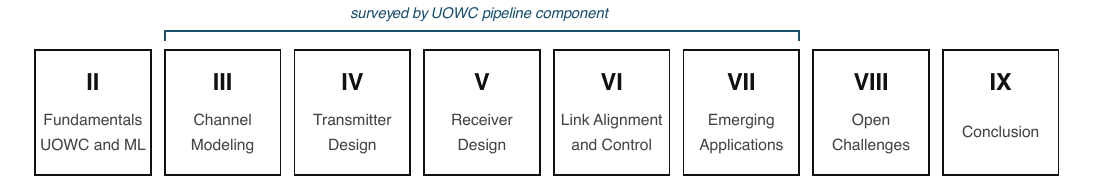}
\caption{Organization of the survey. After the fundamentals of \gls{uowc} and \gls{ml} (Section~\ref{sec:background}), Sections~\ref{sec:channel}--\ref{sec:emerging_applications} survey learning-based methods component by component along the \gls{uowc} pipeline, followed by open challenges and future research directions (Section~\ref{sec:challenges}) and the conclusion (Section~\ref{sec:conclusion}).}
\label{fig:organization}
\end{figure*}
\newcommand{\fullcov}{%
  \tikz[baseline=-0.6ex]\fill (0,0) circle (0.55ex);}

\newcommand{\partcov}{%
  \tikz[baseline=-0.6ex]{
    \draw (0,0) circle (0.55ex);
    \begin{scope}
      \clip (-0.55ex,-0.55ex) rectangle (0,0.55ex);
      \fill (0,0) circle (0.55ex);
    \end{scope}}}

\newcommand{\nocov}{%
  \tikz[baseline=-0.6ex]\draw (0,0) circle (0.55ex);}

\begin{table}[t]
\caption{Comparison of Related Surveys}
\label{tab:related_surveys}
\centering
\footnotesize
\setlength{\tabcolsep}{2pt}
\renewcommand{\arraystretch}{1.15}
\begin{tabular}{@{}
>{\centering\arraybackslash}p{1.0cm}
>{\centering\arraybackslash}p{1.0cm}
>{\centering\arraybackslash}p{0.7cm}
>{\centering\arraybackslash}p{0.7cm}
>{\centering\arraybackslash}p{0.7cm}
>{\centering\arraybackslash}p{0.7cm}
>{\centering\arraybackslash}p{1.0cm}
>{\centering\arraybackslash}p{1.0cm}@{}}
\toprule
\textbf{Year} &
\textbf{Works} &
\textbf{ML} &
\textbf{Ch.} &
\textbf{Tx} &
\textbf{Rx} &
\textbf{Link Align.} &
\textbf{UOWC Apps.} \\
\midrule
2017 & \cite{Zeng2017Survey}
& \nocov & \fullcov & \fullcov & \fullcov & \nocov & \nocov \\

2019 & \cite{Saeed2019}
& \nocov & \fullcov & \fullcov & \fullcov & \nocov & \nocov \\

2023 & \cite{photonics10070811}
& \partcov & \fullcov & \fullcov & \fullcov & \partcov & \nocov \\

2025 & \cite{Li2025RTEsurvey}
& \nocov & \fullcov & \nocov & \nocov & \nocov & \nocov \\

2025 & \cite{ALIMRAN20251026}
& \fullcov & \fullcov & \fullcov & \fullcov & \fullcov & \nocov \\

2025 & \cite{11142888}
& \fullcov & \fullcov & \fullcov & \fullcov & \nocov & \nocov \\

2025 & \cite{11003870}
& \fullcov & \fullcov & \fullcov & \fullcov & \nocov & \nocov \\

2025 & \cite{11053759}
& \fullcov & \nocov & \nocov & \fullcov & \nocov & \nocov \\

2026 & \cite{Shalol2026}
& \fullcov & \fullcov & \fullcov & \fullcov & \nocov & \nocov \\

2026 & \cite{10.1117/12.3108135}
& \fullcov & \partcov & \partcov & \partcov & \nocov & \nocov \\
\midrule
\multicolumn{2}{c}{\textbf{This survey}}
& \fullcov & \fullcov & \fullcov & \fullcov & \fullcov & \fullcov \\
\bottomrule
\multicolumn{8}{l}{\footnotesize
\fullcov~Full coverage;\quad
\partcov~Partial ;\quad
\nocov~Not covered.}\\
\multicolumn{8}{l}{\footnotesize
Ch.: Channel; Tx: Transmitter; Rx: Receiver.}\\
\multicolumn{8}{l}{\footnotesize
Link Align.: Link Alignment; UOWC Apps.: UOWC Applications.}\\
\end{tabular}
\vspace{-2pt}
\end{table}

This paper provides a comprehensive survey of \gls{ml} methods for
\gls{uowc}, organized by pipeline component and supported by a taxonomy
(Table~\ref{tab:taxonomy}) that maps the surveyed studies to their pipeline stages, learning methods, data sources, and evaluation settings.
The survey covers research on the use of \gls{ml} in \gls{uowc} published through May 2026. Its core scope includes underwater optical
channel modeling, transmitter and receiver design, link alignment, and
adaptive control. Closely related studies on underwater imaging, localization, sensing, networking, and power transfer are considered
separately as emerging or adjacent research when their methods support current \gls{uowc} functions or indicate opportunities for future
integration. To achieve the above goals, we first introduce the fundamentals of underwater optical channel physics and provide the basics of various learning algorithms, including \gls{dl} architectures (\gls{mlp}, \gls{cnn}, \gls{rnn}/\gls{lstm}, \gls{ae}, \gls{gan}, \gls{vae}, Transformer, and \gls{dbn}), meta-learning (\gls{maml}), and \gls{rl} (Q-learning, \gls{dqn}, \gls{sarsa}, \gls{ddpg}, and hierarchical and multi-agent \gls{rl}). We link each concept to the \gls{uowc} papers that use it. For each surveyed method, we provide a detailed analysis identifying the specific challenge addressed, the evaluation dataset, and the key results. The survey concludes by identifying and discussing open problems, including the simulation gap, real-time deployment constraints, generalization across water types, system-level optimization, and physics-informed learning.

The remainder of this paper, summarized in Fig.~\ref{fig:organization}, is organized as follows. Section~\ref{sec:background} introduces the fundamentals of \gls{uowc} and \gls{ml}. Sections~\ref{sec:channel}--\ref{sec:emerging_applications} survey learning-based methods by system component: channel modeling and propagation characterization (Section~\ref{sec:channel}), transmitter design and signal processing (Section~\ref{sec:transmitter}), receiver design and signal recovery (Section~\ref{sec:receiver}), link alignment and adaptive control (Section~\ref{sec:link_alignment}), and emerging applications (Section~\ref{sec:emerging_applications}). Section~\ref{sec:challenges} discusses open challenges and future research directions, and Section~\ref{sec:conclusion} concludes the paper.

\begin{table*}[tp]
\caption{Classification of Surveyed Machine Learning Approaches 
According to \acrshort{uowc} System Components}
\label{tab:taxonomy}
\centering
\footnotesize
\setlength{\tabcolsep}{5pt}
\renewcommand{\arraystretch}{1.1}
\begin{tabular}{@{}
>{\raggedright\arraybackslash}p{2.2cm}
>{\raggedright\arraybackslash}p{3.4cm}
>{\raggedright\arraybackslash}p{3.0cm}
>{\raggedright\arraybackslash}p{4.2cm}
>{\raggedright\arraybackslash}p{3.0cm}@{}}
\toprule
\textbf{Pipeline stage} & \textbf{Reference} & 
\textbf{\acrshort{ml} family} & \textbf{Data source} & \textbf{Evaluation} \\
\midrule

\multirow{11}{*}{\parbox{1.8cm}{\textbf{Channel}\\modeling \&\\estimation}}
  & Zhao \textit{et al.}~\cite{9862688}
    & \acrshort{cnn} (AlexNet)
    & Experimental \acrshort{qam} constellation diagrams
    & Experiment \\

  & Al-Amodi \textit{et al.}~\cite{10050890}
    & \acrshort{cnn}
    & Simulated received signal samples 
    & Simulation \\
    & Shashikanth \textit{et al.}~\cite{11478978}
    & \acrshort{cnn}-\acrshort{lstm}
    & Bubble turbulence (15 classes)
    & Experiment \\

  & Lu \textit{et al.}~\cite{9302692}
    & \acrshort{cnn} + \acrshort{dnn}
    & Harbor, clear, and mixed water channels

    & Simulation and experiment \\

  & Jia \textit{et al.}~\cite{11370889}
    & \acrshort{srcnn} + \acrshort{dncnn}
    & Experimental \acrshort{ofdm} FD-\acrshort{uowc} channels
    & Experiment \\
    & Bo \textit{et al.}~\cite{10653174}
    & \acrshort{cnn}-\acrshort{ae} hybrid & \acrshort{egg} turbulence model & Simulation \\
  & Wang \textit{et al.}~\cite{11027977}
    & \acrshort{maml}-\acrshort{dcgan} & 2.4\,m \acrshort{uowc} link &  Experiment \\
& Huo \textit{et al.}~\cite{10.3389/fmars.2023.1149895}
    & \acrshort{dccgan} & 35\,m \acrshort{uowc} link (3 water types) & Experiment \\
    & Du \textit{et al.}~\cite{du2022partially}
    & \acrshort{ppdnn} 
    & \acrshort{mc}-simulated \acrshort{uowc} 
    & Simulation \\

  & Xu \textit{et al.}~\cite{11271218}
    & \acrshort{cnn}-FCSM & Argo ocean data
& Simulation \\
    & Cai \textit{et al.}~\cite{Cai2023Ensemble}
    & Ensemble (Linear + \acrshort{ssbi} + ReLU)
    & Experimental \acrshort{vlc} (1.2\,m tank)
    & Experiment \\
\midrule

\multirow{15}{*}{\parbox{2.1cm}{\textbf{Transmitter}\\design \&\\modulation}}
  & Zhai~\cite{10.1145/3398329.3398367}
    & \acrshort{ae}  & Attenuation + AWGN channel & Simulation \\
  & Zou \textit{et al.}~\cite{9514508}
    &  Multicarrier \acrshort{ae} & \acrshort{mc} impulse responses & Simulation \\
  & Jin \textit{et al.}~\cite{jin2024neural}
    & 2D-AOAE (\acrshort{cnn} + \acrshort{mlp})
    & 1.2\,m \acrshort{uvlc}, green \acrshort{led}
    & Experiment  \\
    & Hu \textit{et al.}~\cite{10819464}
    & Adversarial \acrshort{ae}
    & Water-to-air covert optical link
    & Simulation and experiment \\
    & Zou \textit{et al.}~\cite{10044696}
    & \acrshort{ae}-\acrshort{gan} & Simulated coastal \acrshort{uowc}  & Simulation \\
  & Chauhan \textit{et al.}~\cite{10433978}
    & \acrshort{cnn} (SqueezeNet) & Gamma-Gamma fading & Simulation \\
  & Bisla \textit{et al.}~\cite{10931961}
    & \acrshort{cnn} (SqueezeNet) & Constellation images (OptiSystem) & Simulation \\
  & Singh~\cite{11310834}
    & PML-\acrshort{nn}-AM & Synthetic \acrshort{uowc} dataset& Simulation \\
  & Zhao \textit{et al.}~\cite{10.1016/j.comnet.2024.110233}
    & SwitchOpt \acrshort{rnn} & \acrshort{ofdm} waveforms & Simulation \\
    & Xu \textit{et al.}~\cite{Xu:24}
    & ResNet + \acrshort{lstm}
    & EUVP
    & Emulated \acrshort{uowc} \\
  & Xu \textit{et al.}~\cite{10729883}
    & Swin Transformer  & EUVP, UIEB, RUIE & Emulated \acrshort{uowc} \\
  & Nennouche \textit{et al.}~\cite{11288833}
    & \acrshort{vae} & EUVP, UIEB, LSUI & Simulation \\
      & Xu \textit{et al.}~\cite{11389779}
    & Swin Transformer
    & EUVP, UIEB, RUIE 
    & Emulated \acrshort{uowc}  \\

    & Hu \textit{et al.}~\cite{11005395}
& ConvNeXt
& CIFAR-10, UFO-120
& Simulation\\
  & Lin \textit{et al.}~\cite{Lin:26}
    & \acrshort{rq}-\acrshort{vae} & EUVP &  Experiment \\
\midrule

\multirow{15}{*}{\parbox{2.1cm}{\textbf{Receiver}\\detection,\\demodulation \&\\equalization}}
  & Jiang \textit{et al.}~\cite{8962099}
    & Two-\acrshort{mlp}  & \acrshort{mc}-simulated \acrshort{uowc}  & Simulation \\
  & Amran \textit{et al.}~\cite{11555415}
    & \acrshort{lstm} 
    & Simulated \acrshort{uvlc}
    & Simulation  \\
  & Wang \textit{et al.}~\cite{9817578}
    & \acrshort{maml} + \acrshort{mlp} & 4 water types & Simulation \\
  & Shuai \textit{et al.}~\cite{10495815}
    & \acrshort{dbn} + AdaBoost & 10 modulation schemes & Experiment \\
  & Ma \textit{et al.}~\cite{9209913}
    & \acrshort{dbn} & 10 modulation schemes & Experiment \\
& Nennouche \textit{et al.}~\cite{10636416}
& \acrshort{knn}
& Experimental modulation dataset
& Experiment \\
& Wang \textit{et al.}~\cite{11109827}
    & Shallow \acrshort{dnn} & Synthetic + real \acrshort{ook} & Simulation and experiment \\
    & Salama \textit{et al.}~\cite{Salama2025LowSNR}
    & \acrshort{tcn}-\acrshort{lstm}-AM
    & \acrshort{pam}, \acrshort{qpsk}-\acrshort{ofdm}
    & Simulation \\
& Huang \textit{et al.}~\cite{Huang2026OAM}
& \acrshort{cnn}
& Simulated \acrshort{oam} multiplexing
& Simulation \\

& Cai \textit{et al.}~\cite{9880586}
& \acrshort{bgru}
& 64-\acrshort{apsk} waveforms
& Experiment \\    & Cui \textit{et al.}~\cite{Cui2019OAM}
    & \acrshort{cnn}
    & \acrshort{oamsk} (1\,m tank)
    & Experiment \\
  & Wang \textit{et al.}~\cite{10089049}
    & \acrshort{deepesn} & \acrshort{pam}4, \acrshort{qam}-\acrshort{ofdm} & Experiment \\
& Yousef \& El-Eraki~\cite{Yousef2026DBN}
    & \acrshort{dbn}
    & 4 water types
    & Simulation \\
& Du \textit{et al.}~\cite{Du2024SemiTNN}
& \acrshort{semitnn}
& \acrshort{pam}4 \acrshort{uowc}
& Experiment \\
\bottomrule
\end{tabular}
\end{table*}

% IEEE continued-table convention: same number, caption repeated with
% (Continued), column headings repeated.
\begin{table*}[tp]
\addtocounter{table}{-1}
\caption{Classification of Surveyed Machine Learning Approaches 
According to \acrshort{uowc} System Components (Continued)}
\centering
\footnotesize
\setlength{\tabcolsep}{5pt}
\renewcommand{\arraystretch}{1.1}
\begin{tabular}{@{}
>{\raggedright\arraybackslash}p{2.2cm}
>{\raggedright\arraybackslash}p{3.4cm}
>{\raggedright\arraybackslash}p{3.0cm}
>{\raggedright\arraybackslash}p{4.2cm}
>{\raggedright\arraybackslash}p{3.0cm}@{}}
\toprule
\textbf{Pipeline stage} & \textbf{Reference} & 
\textbf{\acrshort{ml} family} & \textbf{Data source} & \textbf{Evaluation} \\
\midrule

\multirow{12}{*}{\parbox{2.1cm}{\textbf{Link}\\\textbf{alignment} \&\\adaptive\\control}}
  & Jia \textit{et al.}~\cite{10663260}
    & YOLOv8s + RTMPose-t & ULDB image dataset & Experiment \\
    & Lu \textit{et al.}~\cite{9455389}
& \acrshort{cnn}
& $2\times2$ \acrshort{mimo} \acrshort{uowc}
& Simulation and experiment\\
& Tanaka \textit{et al.}~\cite{10926134}
& \acrshort{cnn}
& Twin-beam images (2.4\,m tank)
& Experiment\\

  & Ishida \textit{et al.}~\cite{11003393}
    & \acrshort{lstm} + \acrshort{dqn} & Wave displacement data & Simulation \\
  & Romdhane \& Kaddoum~\cite{9770197}
    & Q-learning + \acrshort{sarsa} & 4 water types & Simulation \\
  & Shin \textit{et al.}~\cite{11257798}
    & TPTA-\acrshort{drl} (3 agents) & \acrshort{usv} trajectory data & Simulation \\
  & Shin \textit{et al.}~\cite{10215370}
    & Two-phase two-agent \acrshort{drl} & \acrshort{usv} trajectory data & Simulation \\
  & Shin \textit{et al.}~\cite{10537612}
    & TSTA-\acrshort{drl} (2 agents) & \acrshort{usv} trajectory data & Simulation \\
& Kong \textit{et al.}~\cite{kong2024deep}
    & YOLOv5s
    & \acrshort{apt} images (3\,m tank)
    & Experiment 
    \\
    & Li \textit{et al.}~\cite{10753441}
    & \acrshort{ddpg} +TD3
    & Multi-\acrshort{auv} optical communication
    & Simulation \\
    & Weng \textit{et al.}~\cite{WENG2025121047}
& \acrshort{sac}
& Simulated and sea-trial \acrshort{auv} data
 & Simulation and experiment\\

& Weng \textit{et al.}~\cite{weng2022reinforcement}
& \acrshort{sac}
& Simulated and Tri-TON 2 \acrshort{auv} data
& Simulation and experiment \\
\midrule
\multirow{16}{*}{\parbox{2.1cm}{\textbf{Emerging}\\\textbf{applications}\\(\acrshort{uwpt},\\sensing,\\localization)}}

& Khalfet \textit{et al.}~\cite{11493549}
& \acrshort{gan} (\acrshort{ciecl})
& \acrshort{slipt} over lognormal fading
& Simulation \\

& Shin \textit{et al.}~\cite{shin2024adaptive}
& \acrshort{dqn} + \acrshort{ddpg}
& \acrshort{rov}--sensor link model
& Simulation \\

& Zhang \textit{et al.}~\cite{zhang2026p2dnet}
& P$^2$DNet
& NEU-PURGB (1,008 pairs)
& Experiment \\

& Wang \textit{et al.}~\cite{wang2025ladar}
& LADAR
& UIEB, EUVP, LSUI, UIQS
& Experiment \\

& Tan \textit{et al.}~\cite{Tan2026PGMamba}
& PGMamba (\acrshort{sagmamba} + \acrshort{pmgffn})
& UIEB, LSUI
& Experiment \\

& Chen \textit{et al.}~\cite{11185120}
& \acrshort{ann} + transfer learning
& LCC--S \acrshort{uwpt} platform
& Experiment \\

& Krishnan \textit{et al.}~\cite{krishnan2021optical:21}
& \acrshort{cnn}--BiLSTM (3D)
& 630~nm \acrshort{led}, turbid water
& Experiment \\

& Joshi \textit{et al.}~\cite{joshi2024underwater}
& YOLOv4 + \acrshort{cnn}--BiLSTM
& 630~nm \acrshort{led}, turbid water
& Experiment \\

& Wang \textit{et al.}~\cite{wang2023underwater}
& YOLOv7
& URPC (8,199 images)
& Experiment \\

& Yu \textit{et al.}~\cite{yu2025aouod}
& AO-UOD
& Acousto--optic dataset (8,001 image pairs)
& Experiment \\
& Weng \& Maki~\cite{10682234}
& \acrshort{sac}
& Simulated \acrshort{auv} trajectories
& Simulation \\

& Zhang \textit{et al.}~\cite{zhang2025nav}
& SE-ResNet50
& Underwater polarization images
& Experiment \\

& Tiwari \textit{et al.}~\cite{tiwari2025mine}
&  YOLOv8-inspired
& Synthetic mine detection data
& Simulation \\

& Simon \textit{et al.}~\cite{simon2025dual}
& \acrshort{eann} + \acrshort{dqlaecm}
& Aqua-Sim (100 nodes)
& Simulation \\

& Li \textit{et al.}~\cite{li2019multiagent}
& \acrshort{marl}
& UOWSN (24 sensor nodes + 1 sink)
& Simulation \\

& Li \textit{et al.}~\cite{li2020routing}
& DMARL
& UOWSN
& Simulation \\
\bottomrule
\end{tabular}
\end{table*}

% Preserve first-use expansion in the main text for acronyms shown in short
% form in the taxonomy table.
\glsreset{ann}
\glsreset{apsk}
\glsreset{apt}
\glsreset{bgru}
\glsreset{ciecl}
\glsreset{dccgan}
\glsreset{dcgan}
\glsreset{deepesn}
\glsreset{dncnn}
\glsreset{dnn}
\glsreset{dqlaecm}
\glsreset{drl}
\glsreset{eann}
\glsreset{egg}
\glsreset{knn}
\glsreset{marl}
\glsreset{mc}
\glsreset{mimo}
\glsreset{nn}
\glsreset{oam}
\glsreset{oamsk}
\glsreset{ofdm}
\glsreset{ook}
\glsreset{pam}
\glsreset{pmgffn}
\glsreset{ppdnn}
\glsreset{qam}
\glsreset{qpsk}
\glsreset{rov}
\glsreset{rq}
\glsreset{sac}
\glsreset{sagmamba}
\glsreset{semitnn}
\glsreset{slipt}
\glsreset{srcnn}
\glsreset{ssbi}
\glsreset{tcn}
\glsreset{usv}
\glsreset{uvlc}
\glsreset{uwpt}

\section{Fundamentals of UOWC and Machine Learning}
\label{sec:background}
\begin{figure*}[t]
    \centering
    % Source: figures/uwoc-ml-architecture.svg; regenerate the PDF with
    %   rsvg-convert -f pdf -o figures/uwoc-ml-architecture.pdf figures/uwoc-ml-architecture.svg
    \includegraphics[width=\textwidth]{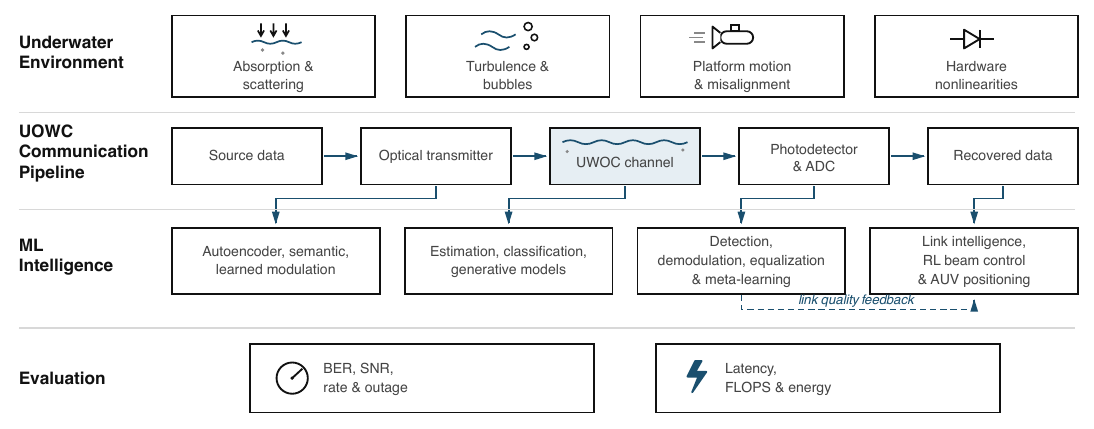}
    \caption{Unified architecture of an ML-enabled UOWC system. Environmental and hardware impairments affect the optical communication pipeline, while learning modules support channel modeling, transceiver optimization, signal recovery, link adaptation, beam control, and AUV positioning.}
    \label{fig:full}
\end{figure*}

This section provides the technical foundation for the survey. It covers both the physical domain and the \gls{ml} tools applied to it. Figure~\ref{fig:full} illustrates the integration of \gls{ml} functions into the \gls{uowc} pipeline. Each \gls{ml} algorithm introduced here is directly linked to the \gls{uowc} problems it addresses in Sections~\ref{sec:channel}--\ref{sec:emerging_applications}. Section~\ref{subsec:bg_channel} describes underwater optical channel physics, including light propagation, \gls{mcrt} simulation, absorption and scattering in seawater, and turbulence and fading models. Section~\ref{subsec:bg_ml} covers \gls{ml} fundamentals. Section~\ref{subsec:bg_dl} introduces the \gls{dl} architectures used throughout the survey. Section~\ref{subsec:bg_meta} introduces meta-learning for environment adaptation. Section~\ref{subsec:bg_rl} presents the \gls{rl} framework used for beam control and power adaptation.

\subsection{Underwater Optical Channel Characteristics}
\label{subsec:bg_channel}
The underwater optical channel is the physical medium through which all \gls{uowc} systems operate. Table~\ref{tab:tech_comparison} compares the three principal underwater wireless communication technologies. Acoustic systems achieve the longest range but are limited to Kbps data rates and suffer from high latency due to the slow propagation speed of sound~\cite{pompili2009overview}. \Gls{rf} systems provide low latency but are severely attenuated by seawater, restricting the useful range to tens of meters. Unlike \gls{rf} and acoustic systems, \Gls{uowc} achieves Gbps data rates and very low latency by operating in the blue-green optical transmission window, but at the cost of limited range and high sensitivity to misalignment~\cite{Kaushal2016}.
\begin{table}[t]
\caption{Comparison of Underwater Wireless Communication Technologies}
\label{tab:tech_comparison}
\centering
\footnotesize
\setlength{\tabcolsep}{4pt}
\renewcommand{\arraystretch}{1.2}
\begin{tabularx}{\columnwidth}{@{}l*{3}{>{\raggedright\arraybackslash}X}@{}}
\toprule
\textbf{Parameter} & \textbf{Acoustic} & \textbf{RF} & \textbf{UOWC} \\
\midrule
Data Rate & $\sim$kbps & $\sim$Mbps & $\sim$Gbps \\
Distance & Up to km & $\leq$10 m & 10--100 m \\
Latency & High & Moderate & Low \\
Frequency Band &
10--15~kHz &
30--300~Hz (ELF) / MHz range (buoyant) &
$10^{12}$--$10^{15}$~Hz \\
Key Limitation & Low data rate, high latency & Severe seawater attenuation & Limited range, misalignment sensitive \\
\bottomrule
\end{tabularx}
\vspace{-2pt}
\begin{flushleft}
\scriptsize Adapted from~\cite{Kaushal2016}
\end{flushleft}
\end{table} 
Many \gls{uowc} researchers utilize \gls{rte} in their channel models. \gls{rte} governs the spatial and angular distribution of radiance $L(\mathbf{r}, \hat{s})$ at position $\mathbf{r}$ in direction $\hat{s}$:
\begin{multline}
    \hat{s} \cdot \nabla L(\mathbf{r}, \hat{s})
    = -c(\mathbf{r})\, L(\mathbf{r}, \hat{s}) \\
    + \int_{4\pi} \beta(\mathbf{r}, \hat{s}' \to \hat{s})\,
      L(\mathbf{r}, \hat{s}')\, \mathrm{d}\hat{s}'
    + S(\mathbf{r}, \hat{s}),
    \label{eq:rte}
\end{multline}
where $c(\mathbf{r}) = a(\mathbf{r}) + b(\mathbf{r})$ is the beam attenuation coefficient, expressed as the sum of the absorption coefficient $a(\mathbf{r})$ and the scattering coefficient $b(\mathbf{r})$; $\beta(\mathbf{r}, \hat{s}' \to \hat{s})$ is the volume scattering function describing the angular redistribution of scattered light; and $S(\mathbf{r}, \hat{s})$ is a source term~\cite{Mobley1994}. The \gls{rte} generally has no closed-form solution due to its mathematical difficulty, which motivates numerical solution methods.

The simplest approximation to the \gls{rte} is the Beer–Lambert law, which models light attenuation along a straight path of length $z$ as:
\begin{equation}
I(z)=I_0 e^{-c(\lambda)z}
\label{eq:beer_lambert}
\end{equation}
where $I_0$ and $I(z)$ represent the transmitted and received optical intensities, respectively, and $c(\lambda)$ is the extinction coefficient~\cite{beer}. Although computationally efficient, the Beer–Lambert law assumes straight-line propagation and neglects multiple scattering and turbulence effects, which may result in inaccurate channel characterization in realistic underwater environments~\cite{Li2025RTEsurvey, Mobley:93}. When multiple scattering or turbulence dominate, the Beer–Lambert attenuation alone may no longer provide an adequate channel description. 
\Gls{mcrt} is the most widely used numerical method for this purpose in \gls{uowc} channel simulation. The \gls{mcrt} method simulates the propagation of a large number of photon packets through the underwater medium, modeling each absorption and scattering event randomly using the \gls{iop} of water. For each photon packet, the free path length $\ell$ between successive interaction events is drawn from an exponential distribution:
\begin{equation}
    \ell = -\frac{\ln(u)}{c(\lambda)},
    \label{eq:mcpath}
\end{equation}
where $u \sim \mathcal{U}(0,1)$ is a uniform random 
variable. At each interaction point, a Bernoulli trial 
with probability $\omega_0 = b/c$ determines whether 
the event is a scattering or absorption event~\cite{mobley2022oceanic,Mobley1994}. If 
scattered, the new photon direction is sampled from 
the volume scattering function. The channel impulse 
response is estimated by accumulating the weights of 
photon packets that reach the 
receiver~\cite{Tang2014Impulse}. Although \gls{mcrt} 
requires a large number of photons for low noise 
estimates, it is physically accurate. Recent work has 
extended the \gls{mcrt} framework to jointly model 
absorption, scattering, and turbulence using multiple 
phase screens~\cite{Wen2023MCMPS}.

The characteristics of the underwater optical channel depend strongly on the surrounding environment. In particular, absorption and scattering properties vary significantly across different water types. Pure seawater exhibits minimum attenuation in the blue–green spectral region, which defines the optimal transmission window for \gls{uowc} systems. Four canonical water types are commonly adopted in the \gls{uowc} literature as a benchmark for channel models and \gls{ml} algorithms: 
pure seawater, with absorption dominating and the 
longest transmission ranges; clear ocean water, where 
dissolved organic matter increases absorption; coastal 
ocean water, characterized by elevated phytoplankton 
and detritus; and turbid harbor water, the most 
challenging condition for \gls{uowc} design~\cite{Kaushal2016}.
\Gls{uowc} channels are subject to random irradiance fluctuations caused by oceanic turbulence arising from refractive-index perturbations induced by temperature gradients, salinity variations, and air bubbles~\cite{8606206, Ata2023, Xu2023WGG}.

Many statistical distributions have been widely used to model the fading caused by the varying environment. The log-normal distribution is commonly used to model turbulence-induced fading under weak turbulence conditions \cite{andrews2005laser}. The Gamma-Gamma distribution models the combined effects of small- and large-scale irradiance fluctuations and has also been applied to UOWC fading, although its accuracy can vary with the underwater channel conditions~\cite{8370053,10666711,10.1117/1.1386641}. Additionally, the \gls{egg} distribution~\cite{8606206} is an experimentally validated model widely adopted for \gls{uowc} turbulence involving air bubbles and temperature gradients. The \gls{egg} model is adopted in several surveyed papers on channel estimation~\cite{10050890,10653174}. More recently, the \gls{wgg} distribution has been proposed to characterize turbulence-induced fading in vertical \gls{uowc} links, where temperature and salinity vary with depth~\cite{Xu2023WGG}.

\subsection{Machine Learning Fundamentals}
\label{subsec:bg_ml}

\gls{ml} is a subfield of artificial intelligence concerned with algorithms that improve their performance on a task through experience \cite{Goodfellow2016}. \gls{ml} methods learn mappings directly from data by optimizing a parameterized function with respect to a defined objective. In the context of \gls{uowc}, the dynamic nature of the underwater optical channel and the complex, nonlinear relationships between physical channel parameters and system performance can challenge conventional analytical approaches, motivating the use of \gls{ml} techniques~\cite{11310834,OShea2017,Qin2019,He2019ModelDriven}.
\Gls{ml} problems are broadly classified into three categories based on the availability of training data and the learning objective. Supervised learning is the most widely adopted framework in \gls{uowc} applications. It operates on labeled datasets consisting of input observations (e.g., received signal vectors or constellation diagram images) and corresponding target labels (e.g., modulation type or \gls{snr} level). In \gls{uowc}, supervised learning has been utilized in \gls{snr} estimation \cite{9862688}, modulation format classification \cite{10433978, 10931961}, channel parameter estimation \cite{10050890}, and \gls{ber}-based adaptive modulation \cite{11310834}.

Unsupervised learning operates without target labels, seeking instead to discover the underlying distribution of the input data. Common objectives include density estimation, clustering, and dimensionality reduction. In \gls{uowc}, unsupervised pre-training is used in \gls{dbn}-based demodulators \cite{10495815, 9209913, Yousef2026DBN}, where each \gls{rbm} layer learns latent feature representations from received signals without requiring labeled examples, reducing the annotation burden in experimental setups.

Self-supervised learning generates supervisory signals automatically from the input data. The \gls{ae} framework \cite{OShea2017} illustrates this paradigm in communications: the transmitter and receiver are jointly trained to reconstruct the transmitted message, with no external labels required.

These three paradigms span the applications of \gls{ml} in \gls{uowc}. Several classical \gls{ml} architectures have been explored in this survey and evaluated using the metrics defined below.

\subsubsection{Classical Machine Learning Methods}
\label{subsubsec:bg_classical}

 Several classical \gls{ml} algorithms were applied to \gls{uowc} signal processing and continue to serve as baselines against \gls{dl} approaches.

 \glspl{svm}~\cite{Cortes1995SVM} find the maximum-margin hyperplane that separates classes in a feature space. \glspl{svm} served as a benchmark architecture for modulation classification and signal demodulation in \gls{uowc} experiments \cite{10495815}.

\gls{knn} is a non-parametric classification algorithm that assigns a query sample to a class based on the majority vote among its $k$ nearest neighbors in the feature space \cite{1053964}. In \gls{uowc} systems, \gls{knn} has been investigated for signal classification and demodulation \cite{10495815}. It has also been evaluated as a demodulator across multiple modulation schemes~\cite{10636416}.

Ensemble methods such as AdaBoost combine multiple weak classifiers through adaptive sample reweighting~\cite{FREUND1997119}. AdaBoost-based demodulators using cascaded \gls{knn} weak classifiers have been shown to outperform both \gls{svm} and naive Bayes classifiers on experimental \gls{uowc} datasets, demonstrating the value of ensemble approaches for physical-layer signal classification \cite{10495815}.

\subsubsection{Evaluation Metrics}
\label{subsubsec:bg_metrics}

The metrics used throughout this survey are defined here for reference.

The \textit{\gls{ber}} measures the fraction of incorrectly decoded bits and is the primary reliability metric for digital \gls{uowc} links~\cite{proakis2001digital}. It is plotted as a function of received \gls{snr} to characterize system performance across channel conditions.

The \textit{\gls{snr}} quantifies the ratio of signal power $P_s$ to noise power $P_n$:
\begin{equation}
    \text{SNR} = 10\log_{10}\!\left(\frac{P_s}{P_n}\right) \;\text{[dB]}.
\end{equation}

The \textit{\gls{psnr}} and \textit{\gls{ssim}} \cite{1284395} are image quality metrics used to evaluate semantic communication and underwater image enhancement systems. \gls{psnr} measures the ratio of peak signal power to reconstruction error in decibels, while \gls{ssim} captures perceptual similarity in terms of luminance, contrast, and structural information.
\textit{Classification accuracy} is used for modulation recognition and water type classification tasks, defined as the fraction of correctly classified samples over the full test set. \textit{\gls{nmse}} is used for channel estimation and parameter regression tasks.

\subsection{Deep Learning Architectures}
\label{subsec:bg_dl}

\Gls{dl} refers to a class of \gls{ml} methods built on multiple layered \glspl{ann} that learn hierarchical representations directly from raw inputs \cite{LeCun2015, Goodfellow2016}. This subsection introduces the principal architectures employed throughout this survey.
\subsubsection{Feedforward Neural Networks}
\label{subsubsec:bg_fnn}

The feedforward neural network, or \gls{mlp}, is the foundational building block of \gls{dl}. An \gls{mlp} consists of an input layer, one or more hidden layers, and an output layer, where each layer applies a learned affine transformation followed by a nonlinear activation function. Given an input vector $\mathbf{x} \in \mathbb{R}^{d}$, the output of the $l$-th hidden layer is computed as:
\begin{equation}
    \mathbf{h}^{(l)} = \sigma\!\left(\mathbf{W}^{(l)}\mathbf{h}^{(l-1)} + \mathbf{b}^{(l)}\right),
\end{equation}
where $\mathbf{W}^{(l)}$ and $\mathbf{b}^{(l)}$ are the learnable weight matrix and bias vector of layer $l$, and $\sigma(\cdot)$ denotes the activation function. In \gls{uowc}, \glspl{mlp} have been employed for channel compensation, equalization, and signal demodulation \cite{8962099,9817578}.

\subsubsection{Convolutional Neural Networks}
\label{subsubsec:bg_cnn}

\Glspl{cnn} extend the \gls{mlp} by introducing shared convolutional filters that exploit spatial or temporal locality in the input \cite{Krizhevsky2012}. A convolutional layer applies a set of learned filters across the input feature map, producing output feature maps that encode local patterns.

\begin{figure}[t]
    \centering
    % Source: figures/uwoc-cnn-receiver.svg; regenerate the PDF with
    %   rsvg-convert -f pdf -o figures/uwoc-cnn-receiver.pdf figures/uwoc-cnn-receiver.svg
    \includegraphics[width=\columnwidth]{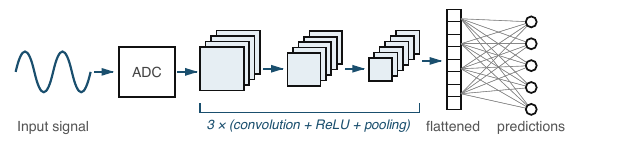}
    \caption{\Gls{cnn} architecture operating on digitized received signal samples. }
    \label{fig:cnn}
\end{figure}

Convolutional layers are typically followed by pooling layers that reduce spatial dimensionality, as illustrated in Fig. \ref{fig:cnn}. This hierarchical structure enables the network to learn increasingly abstract feature representations, from low-level edges and textures in early layers to high-level semantic patterns in deeper layers. In \gls{uowc}, \glspl{cnn} have been applied to \gls{snr} estimation~\cite{9862688}, modulation classification \cite{10433978, 10931961}, channel parameter estimation \cite{10050890}, water-type classification \cite{9302692}, fast channel simulation \cite{11271218}, \gls{oam} signal decoding \cite{Cui2019OAM, Huang2026OAM}, and misalignment estimation for link alignment \cite{9455389, 10926134}.

\subsubsection{Recurrent Neural Networks and LSTM}
\label{subsubsec:bg_rnn}

\Glspl{rnn} are neural networks designed for sequential data. By maintaining an internal hidden state, they can capture temporal dependencies that are difficult to model using feedforward networks. However, conventional \glspl{rnn} suffer from the vanishing gradient problem when learning long-range dependencies.

To address this limitation, the \gls{lstm} architecture introduces gated memory cells that regulate information flow through forget, input, and output gates \cite{Hochreiter1997LSTM,818041}. The \gls{gru} further simplifies this design by combining certain gating mechanisms, reducing computational complexity while maintaining competitive performance \cite{Cho2014GRU}.

In \gls{uowc}, recurrent architectures have been applied to channel modeling and prediction \cite{Xu:24}, bubble-induced channel classification \cite{11478978}, joint channel estimation and signal detection \cite{11555415}, beam steering \cite{11003393}, and learning-assisted equalization \cite{Salama2025LowSNR,9880586}.

\subsubsection{Autoencoders}
\label{subsubsec:bg_ae}

An \gls{ae} jointly trains an encoder $f_\phi(\cdot)$ and a decoder $g_\theta(\cdot)$ to minimize reconstruction loss \cite{Zhai2018AE}. The encoder compresses the input into a latent representation, while the decoder reconstructs the original input. Training is performed by minimizing the reconstruction loss between the input and reconstructed output.
For \gls{uowc}, \gls{ae}-based transceivers learn channel-adapted constellations that outperform conventional \gls{qam} and \gls{pam} under specific impairment conditions~\cite{10.1145/3398329.3398367,9514508} and \gls{ae} hybrid architectures~\cite{10653174} assist in channel modeling and performance analysis.
% \begin{figure}[t]
%     \centering
%     \includegraphics[width=\columnwidth]{figures/autoencoder.png}
%     \caption{General autoencoder architecture with encoder, latent space, and decoder. }
%     \label{fig:ae}
% \end{figure}

\subsubsection{Generative Adversarial Networks}
\label{subsubsec:bg_gan}

A \gls{gan} \cite{Goodfellow2014GAN} trains a generator $G_\theta(\mathbf{z})$ and a discriminator $D_\phi(\mathbf{x})$ via a minimax objective. At convergence, the generator produces samples indistinguishable from real data. The loss is calculated from both the generator and discriminator, where the discriminator is trained to distinguish real samples from generated ones, while the generator is trained to fool the discriminator by producing increasingly realistic samples.
In \gls{uowc}, \glspl{gan} serve three complementary roles: covert signal generation, where transmitted signals are shaped to resemble noise~\cite{10819464}; channel emulation, where the generator learns the statistical distribution of the underwater channel to augment training datasets~\cite{11027977, 10.3389/fmars.2023.1149895}; and power transfer in \cite{11493549}.
% \begin{figure}[t]
%     \centering
%     \includegraphics[width=\columnwidth]{figures/Gans.png}
%     \caption{\gls{gan} training framework with generator and discriminator networks.}
%     \label{fig:GAN}
% \end{figure}

\subsubsection{Variational Autoencoders}
\label{subsubsec:bg_vae}

A \gls{vae} \cite{Kingma2014VAE} imposes a prior $p(\mathbf{z}) = \mathcal{N}(\mathbf{0}, \mathbf{I})$ on the latent space.
Latent samples are drawn via the reparameterization trick:
$\mathbf{z} = \boldsymbol{\mu} + \boldsymbol{\sigma} \odot
\boldsymbol{\epsilon}$, $\boldsymbol{\epsilon} \sim
\mathcal{N}(\mathbf{0}, \mathbf{I})$. The continuous and regularized latent space makes \glspl{vae} well-suited for semantic \gls{uowc}, where a compressed representation of source data is transmitted and robustly reconstructed at the receiver under turbulence and path loss~\cite{11288833,Lin:26}.
\subsubsection{Transformer}
\label{subsubsec:bg_transformer}

The Transformer~\cite{Vaswani2017} replaces recurrence with scaled dot-product self-attention. Multi-head attention extends this by computing attention over multiple subspaces in parallel. In \gls{uowc}, Transformer-based architectures have been applied to environment semantics-aided communication, where a channel state attention layer adjusts encoding and decoding based on estimated channel conditions~\cite{10729883, 11389779}.
\subsubsection{Deep Belief Networks}
\label{subsubsec:bg_dbn}

A \gls{dbn} \cite{Hinton2006DBN} stacks multiple \glspl{rbm}, each modeled by an energy function. \glspl{dbn} are trained greedily in an unsupervised fashion. This strategy is effective when labeled data is scarce. In \gls{uowc}, \gls{dbn}-based demodulators achieve high accuracy on 16-\gls{qam} and \gls{dcoofdm} signals in real underwater tank experiments \cite{10495815, 9209913}, and have also been applied to signal reconstruction and denoising~\cite{Yousef2026DBN}.

\subsection{Meta Learning}
\label{subsec:bg_meta}
Meta-learning is a subfield of \gls{ml} concerned with
learning-to-learn: rather than optimizing a model for a single task, it optimizes for the ability to adapt quickly to new tasks.
A significant challenge in \gls{uowc} is the mismatch 
between training and deployment environments. Water types, 
turbulence intensities, and link geometries vary 
continuously, while collecting labeled data per scenario 
is costly. Meta-learning addresses this by training models 
to produce initial parameters from which rapid fine-tuning 
with minimal new data is possible~\cite{pmlr-v70-finn17a}.

The \gls{maml} algorithm~\cite{pmlr-v70-finn17a} learns 
initial parameters $\theta$ from which a small number of 
gradient steps on any new task, drawn from a given task 
distribution, yields strong generalization. It consists 
of two nested loops: the inner loop computes 
gradient updates on a small support set for a specific task to produce adapted parameters for each task $i$, while the outer loop refines the shared initialization by minimizing the sum of losses evaluated at the adapted parameters across 
all tasks. \gls{maml} is widely recognized for its ability to generalize across diverse environments while remaining agnostic to the model architecture~\cite{pmlr-v70-finn17a, modelagnosticmeta}. In \gls{uowc}, \gls{maml} has been applied to train \gls{mlp}-based receivers across pure seawater, clear ocean, coastal, and turbid harbor water tasks~\cite{9817578} and generate received signals across varying underwater attenuation conditions~\cite{11027977}.

\subsection{Reinforcement Learning}
\label{subsec:bg_rl}

\Gls{rl} is a learning framework in which an agent interacts with an environment to maximize a long-term cumulative reward \cite{Sutton2018}. At each discrete time step $t$, the agent observes a state $s_t \in \mathcal{S}$, selects an action $a_t \in \mathcal{A}$ according to a policy $\pi(a_t|s_t)$, receives a scalar reward $r_t$, and transitions to a new state $s_{t+1}$. The goal is to find the optimal policy $\pi^*$ that maximizes the expected discounted return~\cite{li2017deep}.

\begin{figure}[t]
    \centering
    % Source: figures/uwoc-rl-loop.svg; regenerate the PDF with
    %   rsvg-convert -f pdf -o figures/uwoc-rl-loop.pdf figures/uwoc-rl-loop.svg
    \includegraphics[width=\columnwidth]{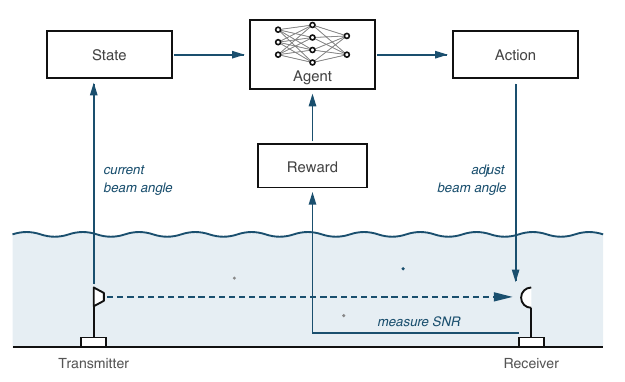}
    \caption{\gls{rl} agent-environment interaction loop for \gls{uowc} beam adaptation.}
    \label{fig:rl}
\end{figure}

In \gls{uowc}, the state typically encodes the current beam parameters, as shown in Fig.~\ref{fig:rl}, or link-quality indicators; the action adjusts beam divergence, orientation, or transmission power; and the reward is a function of the received \gls{snr} or link-maintenance probability~\cite{9770197, 11003393}.

\subsubsection{Q-Learning}
\label{subsubsec:bg_dqn}

Q-learning \cite{Sutton2018} is an off-policy \gls{rl} algorithm that learns the action-value function $Q^*(s,a)$, representing the expected return from taking action $a$ in state $s$ and following the optimal policy thereafter. The update rule is:
\begin{equation}
\begin{aligned}
Q(s_t,a_t) \leftarrow Q(s_t,a_t)
&+ \alpha_Q \Big[
r_t + \gamma \max_{a'} Q(s_{t+1},a')  \\
&\quad - Q(s_t,a_t)
\Big],
\end{aligned}
\label{eq:qlearn}
\end{equation}
where $\alpha_Q$ is the learning rate. The \gls{dqn} \cite{Mnih2015DQN} replaces the tabular $Q$-function with a \gls{dnn} $Q(s,a;\boldsymbol{\theta})$ parameterized by $\boldsymbol{\theta}$, enabling generalization to high-dimensional spaces. In \gls{uowc}, \gls{dqn} is used for beam divergence and orientation control \cite{10215370, 10537612, 11003393}.

\subsubsection{SARSA}
\label{subsubsec:bg_sarsa}

\gls{sarsa} \cite{Sutton2018,rummery1994line} is an on-policy \gls{rl} algorithm that updates the Q-function using the action actually taken by the current policy, rather than the greedy action.
Because \gls{sarsa} learns the value of the current exploration policy rather than the optimal policy, it tends to converge more cautiously than Q-learning in stochastic environments. In \gls{uowc} beam adaptation, \gls{sarsa} has been shown to converge faster than Q-learning for beamwidth and orientation adjustment tasks, achieving approximately 93\% optimal beamwidth selection after 100 iterations \cite{9770197}.

 \subsubsection{Policy Gradient Methods}
\label{subsubsec:bg_ddpg}

\gls{dqn} and Q-learning are limited to discrete action spaces. For \gls{uowc} problems with continuous control variables such as beam divergence angles, the \gls{ddpg} algorithm \cite{Lillicrap2016DDPG} extends the actor-critic framework to continuous action spaces. \gls{ddpg} maintains two networks: a deterministic policy (actor) $\mu(s;\boldsymbol{\theta}^\mu)$ that maps states directly to actions, and a critic $Q(s,a;\boldsymbol{\theta}^Q)$ that evaluates action quality. The actor is updated by ascending the gradient of the critic's Q-value with respect to the actor parameters.
In \gls{uowc}, \gls{ddpg} is used by the sensor-side agent in
hierarchical \gls{slipt} control to continuously optimize
time-switching and power-splitting ratios for energy
harvesting~\cite{shin2024adaptive}, and for collaborative movement control in multi-\gls{auv} optical links~\cite{10753441}.
\subsubsection{\gls{sac}}
\label{subsubsec:sac}

\gls{sac}~\cite{sac} is an off-policy actor-critic algorithm that augments the standard reward objective with an entropy
regularization term, encouraging the agent to maximize both
cumulative reward and the randomness of its policy simultaneously.
The entropy term encourages exploration and can improve training stability relative to deterministic actor–critic methods. However, the resulting performance remains sensitive to reward design, state representation, hyperparameters, and the dynamics of the considered \gls{uowc} environment. In \gls{uowc}, \gls{sac} has been applied to link alignment and
beam control~\cite{weng2022reinforcement,WENG2025121047} and to
autonomous underwater vehicle navigation~\cite{10682234},
demonstrating stable convergence under continuous action spaces
and dynamic channel conditions.

\subsubsection{Hierarchical and Multi-Agent Reinforcement Learning}
\label{subsubsec:bg_marl}

Many \gls{uowc} control problems involve multiple interdependent decisions operating at different time scales. \textit{Hierarchical \gls{rl}} decomposes such decision-making across different levels of temporal abstraction, where higher-level policies can select subgoals or temporally extended actions that are executed by lower-level policies \cite{Sutton2018}. In \gls{uowc}, hierarchical \gls{drl} frameworks have been proposed for joint beam divergence and power control \cite{10215370}, and for simultaneous optical power delivery and data transmission in \gls{slipt} systems \cite{shin2024adaptive}.

\gls{marl} addresses settings in which multiple agents share the environment. In \gls{uowc}, the Three-Agent \gls{drl} (TPTA-DRL) framework deploys three cooperative agents for simultaneous beam orientation, beam divergence, and transmission power control \cite{11257798}.

The \gls{ml} techniques introduced in this section will be discussed within their respective \gls{uowc} system components. The surveyed works are summarized in Table~\ref{tab:taxonomy} and analyzed to demonstrate how different learning frameworks can address existing challenges and limitations in the \gls{uowc} literature.

\section{Channel Modeling and Propagation Characterization}
\label{sec:channel}
Given the \gls{ml} and \gls{uowc} fundamentals 
presented in Section~\ref{sec:background}, this section 
reviews \gls{ml} and \gls{dl} methods applied to 
underwater optical channel modeling and parameter 
estimation. The first part covers approaches for estimating channel parameters and environmental states directly from received signals, including \gls{snr} estimation and channel classification. The second part reviews channel modeling methods, ranging from discriminative emulators to generative 
and physics-informed models.

\subsection{Channel Parameter Estimation}

Accurate characterization of the dynamic underwater channel is crucial for optimizing the performance of \gls{uowc} systems. However, conventional analytical approaches often struggle to capture the highly nonlinear and stochastic relationships between environmental conditions and channel behavior. \Gls{ml} methods can learn complex mappings between received signals, channel characteristics, and environmental conditions, enabling direct estimation of channel parameters and channel states from data.

Zhao \textit{et al.}~\cite{9862688} formulated \gls{snr}
estimation as a constellation-image classification problem.
Received \gls{ofdm} symbols from a 2-m experimental \gls{uowc}
link were converted into \(224\times224\) constellation diagrams and used as inputs to an AlexNet-based classifier. The final output layer was modified from 1000 to 13 classes, corresponding to 13 discrete \gls{snr} values. For each modulation format, the dataset contained 1,560 constellation diagrams. The proposed method achieved \gls{snr}-classification accuracies of 99.7\%, 98\%, and 94.7\% for 2-\gls{qam}, 4-\gls{qam}, and 8-\gls{qam}, respectively.

Beyond \gls{snr} estimation, predicting the full channel state enables joint optimization of classification, estimation, and detection. An adaptive communication system that jointly performs these tasks was proposed in \cite{9302692}. The architecture consists of a one-dimensional \gls{cnn} for channel classification and a \gls{dnn} that jointly performs channel estimation and signal detection. The channel classifier identifies the water type and produces estimated combinational weights that are used to combine the outputs of multiple \gls{dnn} detectors trained for different water environments, allowing the receiver to adapt to varying channel conditions. The system was trained using Monte Carlo generated \gls{uowc} channels representing harbor, clear, and mixed water types and was validated through both simulations and water tank experiments. Results show significant \gls{ber} improvements over conventional receivers. 

Another application for channel classification was proposed in~\cite{11478978}. The authors experimentally analyzed 15 types of air-bubble attenuation using a 530~nm \gls{led}-based \gls{imdd} optical link. The experimental setup included three bubble sizes (small, medium, large) at varying pump rates and two beam widths, generating diverse fading conditions. Statistical analysis revealed that different bubble conditions exhibit distinct distributions, with the \gls{wgg} applicable for large bubbles, lognormal or Normal for medium bubbles, and Normal or Weibull for small bubbles, with scintillation indices spanning two orders of magnitude. Coherence time was observed to decrease with bubble size and beam width. A \gls{cnn}-\gls{lstm} neural network was designed to classify the bubble attenuation type from pilot symbols (5\% overhead), achieving 92.27\% accuracy and an 89\% F1 score across all 15 classes. This work demonstrates the effectiveness of
\gls{cnn}-\gls{lstm} architectures for channel classification
in bubble-induced underwater turbulence.

For parameter estimation in the channel model, the authors in~\cite{10050890} developed a \gls{dl} approach for \gls{uowc} channel characterization using the \gls{egg} channel model. The network estimates the \gls{egg} parameters directly from received signal samples collected under different bubble levels and temperature gradient conditions, while a second \gls{nn} maps the estimated parameters to the corresponding environmental states. The model was trained on signals generated from 16 \gls{egg} channel scenarios derived from experimentally measured channel parameters. Simulation results demonstrated \gls{nmse} ranging from
approximately \(10^{-4}\) to \(10^{-3}\) for \gls{egg}
parameter estimation, with validation \gls{nmse} values of
\(0.001077\) and \(0.004784\) for bubble level and temperature
gradient estimation, respectively.

More recently, \gls{dl} has been applied directly to channel estimation in full-duplex \gls{uowc} systems. Unlike conventional half-duplex links, full-duplex communication enables simultaneous transmission and reception but suffers from severe backscatter interference~\cite{Liu:24}. To address this challenge, Jia \textit{et al.}~\cite{11370889} proposed a cascaded deep convolutional \gls{nn} architecture consisting of a \gls{srcnn} and a \gls{dncnn}. The \gls{srcnn} reconstructs high-resolution channel responses from sparse pilot-based estimates, while the \gls{dncnn} suppresses residual noise and backscatter interference. The proposed framework was experimentally validated on a real full-duplex \gls{uowc} platform with a 1.3\,m turbid-water link. Experimental results demonstrated that the method effectively mitigates backscatter interference, reducing \gls{nmse} by up to 1.93\,dB and improving \gls{ber} performance compared with conventional least-squares estimation.

In addition to data-driven approaches, physics-informed methods embed domain knowledge directly into the network architecture. Cai \textit{et al.}~\cite{Cai2023Ensemble} proposed a physics-based ensemble learning framework for underwater visible light channel estimation. The architecture consists of three specialized subnetworks: a linear estimator for \gls{isi}, a \gls{ssbi} transfer network for quadratic distortion arising from square-law detection by the photodetector, and a ReLU network for higher-order nonlinearities from optoelectronic devices. This ensemble approach is designed based on physical prior knowledge: the linear, quadratic, and higher-order subnetworks explicitly model \gls{isi}, \gls{ssbi}, and optoelectronic nonlinearities, respectively. Experimental validation in a 1.2-m water tank demonstrated that the Ensemble estimator achieves significantly lower \gls{mse} than \gls{lms} and single network estimators, while also learning the characteristic V-shaped Vpp-BER curve of the underwater channel. This work demonstrates that combining physical understanding with the nonlinear approximation power of neural networks enables systems to learn channel characteristics beyond what conventional model-based approaches can capture.

The methods in this subsection estimated channel
parameters, predicted channel states, and classified channel
conditions to build physically aware systems and improve
\gls{uowc} performance. The \gls{snr} estimation
method~\cite{9862688} achieved classification accuracies of
99.7\%, 98\%, and 94.7\% for 2-\gls{qam}, 4-\gls{qam},
and 8-\gls{qam} modulation schemes. The channel state
prediction model~\cite{9302692} resulted in lower \gls{ber}
than conventional methods. The bubble attenuation
classifier~\cite{11478978} achieved 92.27\% accuracy and an
89\% F1 score across 15 channel classes. The two-stage
\gls{egg} framework~\cite{10050890} estimated turbulence
parameters with \gls{nmse} values on the order of
\(10^{-4}\) to \(10^{-3}\). Full-duplex channel
estimation~\cite{11370889} reduced channel estimation
\gls{nmse} by up to 1.93\,dB, and physics-guided ensemble
learning~\cite{Cai2023Ensemble} achieved lower \gls{mse}
than \gls{lms} and single network estimators by embedding
physical priors into the network architecture. However, these
methods address distinct channel conditions and are rarely
benchmarked against each other on a common dataset. A
standardized \gls{uowc} channel estimation benchmark covering
multiple water types, turbulence levels, and hardware
configurations remains an important missing contribution.

\subsection{Channel Modeling}

Traditional underwater optical channel models are commonly based on radiative transfer theory, \gls{mc} simulations, or impulse response fitting techniques. While these approaches provide valuable physical insight, they often incur high computational cost and may become difficult to apply. Consequently, recent works have explored data-driven approaches that learn channel behavior directly from measured or simulated data. Early studies demonstrated the feasibility of applying \gls{dl} to communication channel modeling. In acoustic communication, \glspl{dnn} were shown to accurately emulate underwater channels under real water experimental conditions~\cite{onasami2022underwateracousticcommunicationchannel}. 

To address the computational complexity of solving the \gls{rte} for \gls{uowc} path loss prediction, Du \textit{et al.}~\cite{du2022partially} proposed a \gls{ppdnn} combined with parallel \gls{mc} simulation. The parallel \gls{mc} algorithm, employing vectorization and \gls{gpu} acceleration, generates training datasets for three water types (clear, coastal, harbor), with testing distances extended up to 132~m, achieving at least 95\% runtime reduction compared to conventional \gls{mc}. The \gls{dnn}, with three hidden layers containing 10, 20, and 20 neurons, learns the mapping from input parameters to received optical power. A gradual pruning strategy compresses the model by selectively removing low-magnitude weights, reducing storage space with minimal performance degradation. The \gls{ppdnn} outperforms the classical \gls{ml} models: linear regression, support vector regression, and XGBoost, providing instant path loss predictions once trained. The optimal sparsity levels were found to be 0.9, 0.7, and 0.5 for clear, coastal, and harbor waters, respectively, reflecting the increased complexity of fitting scattering-dominated channels.

To capture the nonlinear and time-varying behavior of the underwater optical channel, Bo \textit{et al.}~\cite{10653174} explored a \gls{cnn}-\gls{ae}. The encoder consists of three one-dimensional convolutional layers that progressively extract features from input bit sequences and map them into a low-dimensional space, while the decoder symmetrically reconstructs the original information. The system incorporates \gls{mc} simulations to generate channel impulse responses for different seawater types and employs the \gls{egg} model to account for ocean turbulence effects. The \gls{egg} model demonstrated a goodness-of-fit of 0.9964, with an \gls{rmse} of 0.0048 compared to the dual Gamma and generalized Gamma models. Training data were generated using channel impulse responses obtained from \gls{mc} simulations, while the \gls{egg} model was used to characterize the effects of underwater turbulence.

Building on discriminative emulators, generative approaches
have also been explored to improve generalization across water
types. A proposed method that supports generalization across water types was discussed in~\cite{11027977}. This method combines a \gls{dcgan} with \gls{maml} and treats each water type as a separate learning task. The generator receives the transmitted signal and attenuation coefficient as conditional inputs and synthesizes realistic received signals, while the discriminator distinguishes generated samples from experimental measurements. Using data collected from a 2.4\,m underwater optical link operating under multiple attenuation conditions, the \gls{maml}-\gls{dcgan} achieved a correlation coefficient of 0.902 and a \gls{ber} mismatch of approximately \(10^{-4}\), together with smaller spectrum deviations than conventional \gls{mlp}, \gls{cnn}, and \gls{dcgan}-based emulators.

\begin{figure}[t]
\centering
% Source: figures/uwoc-cgan-emulator.svg; regenerate the PDF with
%   rsvg-convert -f pdf -o figures/uwoc-cgan-emulator.pdf figures/uwoc-cgan-emulator.svg
\includegraphics[width=\columnwidth]{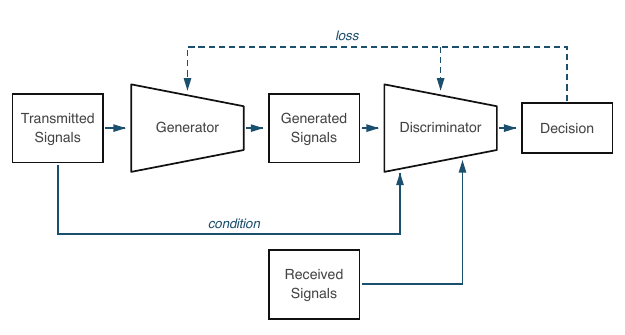}
\caption{\gls{cgan} framework for \gls{uowc} channel emulation: the generator synthesizes received signals conditioned on the transmitted input, while the discriminator distinguishes generated from real received signals.}
\label{fig:cgan}
\end{figure}

Complementing the \gls{dcgan} approach, Huo \textit{et al.} proposed a \gls{dccgan} framework for \gls{uowc} 
channel emulation~\cite{10.3389/fmars.2023.1149895}. 
Unlike standard \glspl{gan}, \glspl{cgan}, as 
illustrated in Fig.~\ref{fig:cgan}, condition both the generator and discriminator on additional information $c$, and in \gls{uowc} channel emulation, $c$ represents the transmitted signal. The \gls{dccgan} combines \gls{cgan} conditioning with deep convolutional 
feature extraction to address a key limitation of 
conventional channel models: their focus on water 
propagation effects while neglecting hardware 
nonlinearities from optoelectronic devices. 
Experimental validation on a 35\,m \gls{uowc} link across three water types demonstrated that the proposed emulator achieved a correlation coefficient of 0.99, significantly outperforming \gls{cnn} and \gls{mlp} baselines, and successfully captured the stochastic 
bimodal distribution of real received signals.

While most existing \gls{uowc} channel models assume static horizontal or vertical links, practical deployments involving mobile platforms introduce slant propagation paths, depth dependent attenuation, and spatially heterogeneous turbulence. To address these challenges, Xu \textit{et al.}~\cite{11271218} developed a unified framework that integrates \gls{wos} with a \gls{cnn}-based Fast Channel Simulation Method (\gls{cnn}-FCSM). The \gls{wos} platform generates physically realistic channel realizations using real ocean measurements, accounting for absorption, turbulence, and platform-induced geometric effects. The \gls{cnn}-FCSM then learns the mapping between environmental parameters and channel responses, serving as a surrogate model for the computationally intensive simulator. Numerical results showed that \gls{cnn}-FCSM achieved more than 98\% prediction accuracy for received optical intensity while reducing computation time by over 90\% compared with conventional \gls{wos}.

Modeling the underwater channel is one of the most complex tasks affecting the overall \gls{uowc} system performance. The parallel \gls{mc} combined with a pruned \gls{dnn}~\cite{du2022partially} achieved at least 95\% runtime reduction compared to conventional
\gls{mc} and outperformed classical \gls{ml} methods. The
\gls{cnn}-\gls{ae} channel emulator~\cite{10653174} incorporated an \gls{egg} turbulence model with a goodness-of-fit of 0.9964, while the generative \gls{maml}-\gls{dcgan}~\cite{11027977} and \gls{dccgan}~\cite{10.3389/fmars.2023.1149895} achieved
correlation coefficients of 0.902 and 0.99, respectively. Lastly, the \gls{cnn}-FCSM~\cite{11271218} achieved more than 98\% prediction accuracy. Notably, the \gls{cnn}-FCSM was evaluated using physics-based \gls{wos} data informed by real Argo ocean measurements, whereas the other surveyed models relied on simulations or controlled laboratory experiments. Despite these advances, none of the surveyed channel models jointly captures absorption, scattering, turbulence, and hardware nonlinearities within a single unified framework. This gap motivates the end-to-end learning approaches discussed in Section~\ref{sec:transmitter}.

\section{Transmitter Design and Signal Processing}
\label{sec:transmitter}
The transmitter is one of the main components of the \gls{uowc} system. It is responsible for converting digital information into modulated signals for propagation through the underwater channel, traditionally relying on fixed modulation schemes such as \gls{ook}, \gls{ppm}, and \gls{ofdm}~\cite{Palaic2024}. Data transmission fundamentally affects the performance of the \gls{uowc} system. However, the dynamic oceanic environment introduces inherent challenges such as misalignment sensitivity, power efficiency constraints, and fading. This section explores how deep learning techniques enable intelligent transmitter architectures that learn optimal signal representations and transmit semantically aware information. 

\subsection{End-to-End Learning for UOWC Transceivers}

\begin{figure}[t]
\centering
% Source: figures/uwoc-ae-transceiver.svg; regenerate the PDF with
%   rsvg-convert -f pdf -o figures/uwoc-ae-transceiver.pdf figures/uwoc-ae-transceiver.svg
\includegraphics[width=\columnwidth]{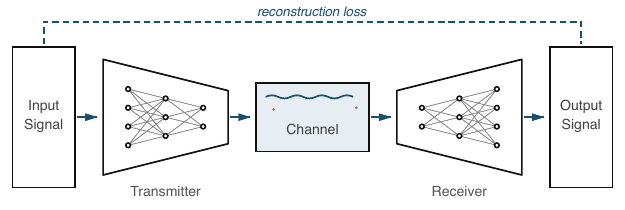}
\caption{\gls{ae} Transceiver Architecture. The transmitter learns a compact signal representation that propagates through the underwater optical channel, while the receiver reconstructs the transmitted information.}
\label{fig:ae_uwoc}
\end{figure}

Unlike conventional communication systems that separately optimize modulation, coding, and detection blocks, end-to-end learning models the transmitter and receiver as the encoder and decoder of an \gls{ae}, as shown in Fig.~\ref{fig:ae_uwoc}, while the channel is represented as an intermediate layer. The entire transceiver is jointly optimized through a common loss function. This enables the system to learn channel-adaptive signal representations directly from data and applies the concept of end-to-end learning first introduced in~\cite{oshea2016learningcommunicatechannelautoencoders}.

The transmission process of the transmitter, channel, and receiver is modeled as a single end-to-end deep neural network trained to minimize a loss function, enabling joint optimization of encoding, modulation, and decoding without relying on separate fixed blocks. In optical fiber communication, the first experimental validation of an end-to-end system represented the entire transceiver using an \gls{ae}~\cite{karanov2018end}. This approach enables the transceiver to learn optimal parameters for both the transmitter and receiver without requiring manual reconfiguration.

\subsubsection{\gls{ae}-Based Transceivers}

\gls{ae}-based transceivers were investigated for \gls{uowc} systems by Zhai~\cite{10.1145/3398329.3398367}, extending the \gls{ae} architecture Zhu~\textit{et al.}~\cite{8819929} proposed for \gls{owc} systems by incorporating underwater attenuation and adapting it to single-channel transmission. Zhai's framework consists of an encoder, a channel layer, and a decoder. The encoder maps digital messages into continuous transmission symbols through fully connected neural network layers, while a constraint layer enforces the non-negativity and peak-power limitations required by \gls{imdd} systems. The channel is modeled using attenuation and additive Gaussian noise, and the decoder estimates the transmitted messages through a softmax classifier. By minimizing the cross-entropy loss during training, the \gls{ae} learns signal constellations that are comparable to or outperform conventional \gls{pam} and \gls{qam} modulation schemes at moderate and high \gls{snr}, with the advantage widening for higher-order \gls{pam}, where conventional methods suffer from performance degradation. Simulations were conducted for both space division multiplexing and single-channel configurations.

Subsequent studies focused on making \gls{ae} architectures more suitable for realistic \gls{uowc} environments. Zou \textit{et al.}~\cite{9514508} extended a single-carrier \gls{ae} framework to a multicarrier architecture by incorporating \gls{dcoofdm}, addressing the severe \gls{isi} encountered in high-data-rate \gls{uowc} systems. The framework jointly optimizes constellation mapping, demapping, and DC bias selection, with a scale factor that is adaptively learned using a loss function combining cross entropy and clipping distortion. Hermitian symmetry is employed to generate real-valued signals, while DC biasing and clipping enforce the non-negativity required by \gls{imdd} systems. \gls{mc}-generated channel impulse responses are regenerated at each training epoch to capture time-varying underwater channel dynamics. These enhancements significantly improve the performance of end-to-end transceivers under realistic propagation conditions, particularly at high data rates where conventional single-carrier \gls{ae} systems become ineffective under severe \gls{isi}.

One of the limitations of traditional \gls{ae} schemes is that they can only generate one-dimensional signals and require manual frequency shifting to avoid the \gls{lfn}. The study in~\cite{jin2024neural} proposed a novel 2D adaptive optimization \gls{ae} (2D-AOAE) framework for \gls{uvlc} systems to address these shortcomings. The framework employs a convolutional transmission network with kernels initialized as Hilbert filter pairs, facilitating automatic \gls{lfn} avoidance and high-frequency compensation while generating orthogonal I/Q data streams. The approach comprises four key components: a mapping network for constellation point generation; a modulation network with convolutional layers for signal shaping and pre-equalization; a dual branch channel model incorporating linear and nonlinear branches with trainable Gaussian noise layers; and demodulation and demapping networks for signal recovery. Experimental validation over a 1.2-m \gls{uvlc} link with a green \gls{led} demonstrates that the model achieves a transmission rate of 2.85\,Gbps, representing a 15.4\% improvement over traditional 32-\gls{qam} \gls{uvlc} systems and a 73\% improvement over conventional \gls{ae} schemes. These results establish that structured architectural constraints can significantly enhance both the performance and efficiency of end-to-end learned transceivers for high speed \gls{uvlc} systems.

\subsubsection{Generative and Adversarial Approaches}

Through adversarial training, the generator progressively learns realistic signal representations that can be exploited for communication tasks such as channel equalization, channel emulation, and covert communications~\cite{8553233}.

Hu \textit{et al.} proposed an adversarial \gls{ae} framework that combines the representation learning capability of \glspl{ae} with the distribution matching property of \glspl{gan}~\cite{10819464}. The transmitter generates random covert signals that embed information while matching the distribution of artificial noise, while the receiver reconstructs the original messages through the decoder. The results demonstrated significantly lower \gls{js} divergence than conventional \glspl{lbc}, indicating improved signal covertness, while maintaining a \gls{bler} of 0.0979 in simulation and 0.0540 in an experimental water-to-air link using a 5\,MHz waveform.

More recently, Zou \textit{et al.} developed a hybrid \gls{ae}-\gls{gan} architecture for \gls{uowc} systems employing \gls{ook} modulation and one-bit quantization~\cite{10044696}. In the proposed framework, the \gls{ae} jointly optimizes the transmitter and receiver, while the \gls{gan} generator acts as a generalized channel equalizer that compensates for underwater channel impairments and quantization distortions. By integrating adversarial learning with end-to-end transceiver optimization, the proposed system achieved superior \gls{ber} performance compared with conventional convolutional code-based and pure \gls{ae} transceiver designs, approaching the lower bound represented by a pure \gls{ae} system with perfect equalization and without quantization. The performance advantage became more pronounced at higher \gls{snr} and spectral efficiency.

End-to-end learning has progressively shifted \gls{uowc} transceiver design. Early \gls{ae}-based transceivers~\cite{10.1145/3398329.3398367,9514508} established the potential of jointly optimizing encoding and decoding, achieving performance comparable to or better than conventional schemes under specific channel and \gls{snr} conditions. Subsequent work extended this concept to multicarrier architectures to mitigate \gls{isi}~\cite{9514508} and to 2D signal generation for automatic \gls{lfn} avoidance, high-frequency compensation, and improved robustness to hardware nonlinearities~\cite{jin2024neural}. Adversarial approaches further broadened the design space beyond \gls{ber} minimization, enabling covert communication~\cite{10819464} and hardware aware equalization under one-bit quantization~\cite{10044696}.

\subsection{Modulation Design and Optimization}
Modulation plays a crucial role in the performance and effectiveness of a \gls{uowc} system. It affects metrics such as \gls{ber} and \gls{snr}, and also impacts power consumption. The choice between different modulation schemes involves a trade-off between energy efficiency, bandwidth, and complexity \cite{zayed2025performance}. Due to the varying environment in \gls{uowc}, modulation format classification and link adaptation are essential for robust communication.

\subsubsection{Modulation Format Classification}

Due to the severe channel impairments in \gls{uowc}, the classification of signal modulation schemes is particularly challenging. Chauhan \textit{et al.}~\cite{10433978} proposed a deep learning algorithm to classify various modulation schemes using a \gls{cnn}. They generated signals based on the Gamma-Gamma fading channel model. The received modulated signals are converted into constellation diagram images to represent the signal distortions introduced by the underwater channel. These images are then fed into a \gls{cnn} based on the SqueezeNet architecture to distinguish among five modulation types: \gls{qpsk}, \gls{8psk}, \gls{4qam}, \gls{16qam}, and \gls{64qam}. The performance of the system is evaluated over a wide range of \gls{snr} values, from $-4$~dB to $12$~dB. The results demonstrate that the classification accuracy strongly depends on the \gls{snr}. As the \gls{snr} increases, the accuracy exceeds $90\%$ for all modulation schemes at $8$~dB, while reaching $100\%$ for \gls{qpsk}. This study provides a proof of concept for using \gls{cnn} classifiers to distinguish among various modulation schemes under noisy underwater channel conditions.

Another study~\cite{10931961} also employed a similar approach for modulation classification, utilizing the same SqueezeNet architecture. However, their work differs in methodology and analysis. Rather than generating the signals directly from a theoretical channel model, the authors produced their dataset through system simulations in \textit{OptiSystem 21}, incorporating Gamma-Gamma fading and creating constellation signal images for the same five modulation types. The dataset generation section reports nine \gls{snr} values spanning from $-2$~dB to $14$~dB, while the reported classification results cover values from $-4$~dB to $12$~dB. The proposed \gls{cnn} model was trained on 1,403 constellation signal images and validated on 602 images. At low \gls{snr} ($-4$~dB), the classification accuracy for 64\gls{qam} was approximately 55\%, while \gls{qpsk} achieved 76\%. At higher \gls{snr} ($12$~dB), \gls{qpsk} reached 100\% accuracy, \gls{8psk} achieved 95\%, 4\gls{qam} 93\%, 16\gls{qam} 90\%, and 64\gls{qam} 91\%. These studies demonstrate the potential of lightweight \glspl{cnn} for modulation classification in \gls{uowc}.

\subsubsection{Modulation Adaptation}

Deep learning has also been applied to assist in modulation adaptation. Singh~\cite{11310834} proposed an adaptive modulation algorithm that employs a physics-guided and \gls{ml}-based neural network for adaptive modulation (PML-NN-AM) to dynamically adjust modulation schemes across varying underwater optical channel conditions. The system integrates a closed-form analytical \gls{ber} expression derived by the author for \(M\)-\gls{qam}, which explicitly models the combined effects of turbulence, absorption, and scattering. This physically informed \gls{ber} model serves as the ground truth for training the neural network, enabling it to predict \gls{ber} performance for multiple modulation orders (4-\gls{qam}, 16-\gls{qam}, 64-\gls{qam}, and 256-\gls{qam}) based on channel parameters such as laser power, transmission range, photodetector responsivity, and receiver noise. In the extended architecture, turbulence strength is included as an additional input feature. The architecture then selects the highest order modulation capable of maintaining a \gls{ber} below the \gls{fec} threshold, thereby maximizing data throughput while preserving link reliability under a coastal water scenario. Simulation results demonstrate modulation switching accuracies of 98.6\% under gradually varying turbulence and 96.3\% under rapid and abrupt turbulence changes.

Additionally, Zhao \textit{et al.}~\cite{10.1016/j.comnet.2024.110233} proposed a system that dynamically selects the optimal channel coding rate and time-frequency spreading configuration, referred to by the authors as the optimal \gls{mcs}. Their system uses only the raw time-domain signal waveform after analog-to-digital conversion to perform this adaptation. This addresses hardware interface constraints that prevent an external processor from accessing the internal channel or \gls{snr} estimates of the receiver. The system utilizes an alternating optimization framework, termed SwitchOpt \gls{rnn}, that tunes and switches among candidate deep recurrent neural network classifiers to learn temporal features from sequential optical signals and select the configuration that maximizes physical-layer throughput. They evaluated the performance of the system using a simulated underwater optical system with \gls{ofdm} modulation, time-frequency spreading, and Turbo coding. The results showed that SwitchOpt \gls{rnn} outperforms conventional \gls{cnn} and \gls{ae}-based classifiers in the low \gls{snr} region, while achieving performance close to a conventional link adaptation method that requires explicit channel and \gls{snr} estimates.

The modulation design methods reviewed in this subsection demonstrate that \gls{ml} can classify modulation schemes and adapt modulation or related link configurations under varying \gls{uowc} channel conditions. Lightweight \gls{cnn} classifiers~\cite{10433978,10931961} achieved up to 100\% accuracy for \gls{qpsk} classification. The physics-guided PML-NN-AM~\cite{11310834} demonstrated modulation switching accuracies of 98.6\% under gradually varying turbulence and 96.3\% under rapid, abrupt turbulence changes. The SwitchOpt RNN~\cite{10.1016/j.comnet.2024.110233} adapted the coding rate and spreading configuration without requiring explicit  \gls{snr} feedback.

\subsection{Semantic-Aware Transmission}

Semantic communication shifts the communication objective from transmitting every bit accurately to conveying the underlying meaning of the information. Instead of optimizing the \gls{ber}, semantic communication systems extract relevant semantic features at the transmitter \cite{9679803}.

\begin{figure*}[!t]
\centering
% Source: figures/uwoc-semantic-transmission.svg; regenerate the PDF with
%   rsvg-convert -f pdf -o figures/uwoc-semantic-transmission.pdf figures/uwoc-semantic-transmission.svg
\includegraphics[width=\textwidth]{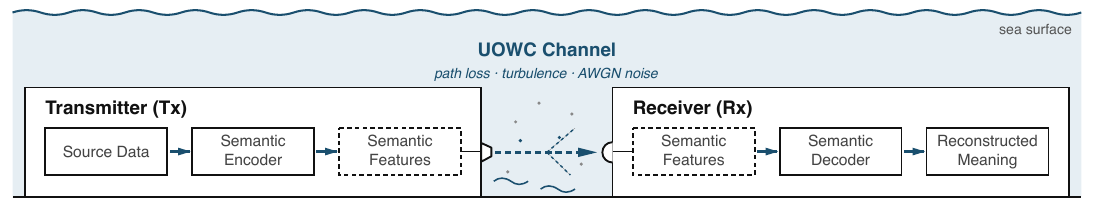}
\caption{Conceptual framework of semantic-aware transmission in
\gls{uowc} systems. Semantic features extracted at the transmitter cross
the underwater optical channel, which introduces path loss, turbulence,
and noise, and are decoded into the reconstructed meaning at the
receiver.}
\label{fig:semantic_uwoc}
\end{figure*}

Xu \textit{et al.}~\cite{Xu:24} introduced a deep learning framework for a \gls{uowc} semantic communication system designed to exploit the semantic information of transmitted signals, as illustrated in Fig.~\ref{fig:semantic_uwoc}. The transceiver is composed of a semantic encoder and a semantic decoder to extract and recover semantic information. Unlike traditional \gls{uowc} systems, their approach employs a deep residual convolutional neural network to extract and transmit semantic image features, thus reducing data redundancy. Overall, the system uses an end-to-end deep learning framework, implementing a semantic-aware transceiver with neural networks and a \gls{lstm} channel model that simulates underwater optical signal propagation during training. The training process comprises two stages: first, the \gls{lstm} channel models are trained under various turbidity conditions to reproduce channel impairments; second, the transceiver is trained while the \gls{lstm} models remain frozen. The system improved performance compared with conventional schemes that rely on \gls{jpeg} compression, \gls{ldpc} coding, and \gls{qam} modulation. Metrics such as \gls{psnr} and \gls{ssim} showed marked improvements, particularly at low bandwidth compression ratios, achieving relative improvements of 26.92\% in \gls{psnr} and 51.90\% in \gls{ssim} over the combined \gls{jpeg}, \gls{ldpc}, and 8-\gls{qam} baseline at a bandwidth compression ratio of 1/16.

An environment semantics aided \gls{uowsc} system was proposed in~\cite{10729883}. The researchers integrated the semantics of the environment into the \gls{uowc} system to estimate the channel state, enabling adaptive control of the transmission code rate and improving bandwidth efficiency. The proposed framework consists of a dark channel prior algorithm that extracts environment semantics related to medium transmittance from a captured image of the channel environment, and a polynomial knowledge base that maps these semantics to the channel gain. It also includes a Swin Transformer based LinkNet for semantic transmission. A channel state attention layer within the Transformer adjusts the encoding and decoding processes based on the estimated channel state information. The final component is a prediction network that predicts the reconstruction distortion for a given semantic feature, scaling parameter, and channel state, allowing the minimum code rate that satisfies the transmission requirement to be selected. The performance was tested using real underwater image datasets (EUVP, UIEB, and RUIE) on an emulated \gls{uowc} system. The system successfully selected code rates that met specified \gls{psnr} thresholds of 24 and 25~dB under varying channel conditions. At an impurity concentration of 8~mg/L, it achieved a \gls{bd}-\gls{bcr} saving of 91.43\% at equivalent \gls{psnr}, together with a \gls{bd}-\gls{psnr} gain of 2.09~dB and a \gls{bd}-\gls{ssim} gain of 0.0864 at the same \gls{bcr}, compared with the baseline \gls{uowsc} scheme.

Following these two fundamental works on semantic underwater communication, Nennouche \textit{et al.}~\cite{11288833} proposed a \gls{vae}-based semantic communication system for \gls{uowc} that transmits a compressed latent representation of an image rather than raw bits. The system employs a pretrained \gls{vae}, where the encoder compresses the input image and the decoder reconstructs it from the noisy received representation. A two-stage training strategy is adopted: noise-free fine-tuning adapts the model to underwater images, while noise-aware fine-tuning improves robustness to channel impairments. The model is trained on the EUVP, UIEB, and LSUI datasets and evaluated using a theoretical channel model that accounts for path loss and oceanic turbulence. Compared with a conventional \gls{jpeg}, \gls{ldpc}, and \gls{ook} system, the proposed model avoids the waterfall effect and achieves a \gls{psnr} of 22~dB and a \gls{ssim} of 0.68 at 95~m in clear water, and a \gls{psnr} of 22.57~dB and a \gls{ssim} of 0.609 at 70~m in coastal water.

Building on this variational approach, Lin \textit{et al.}~\cite{Lin:26} proposed a \gls{rq}-\gls{vae}-based semantic communication system integrated with N-D CAP modulation. A semantic encoder extracts compact latent representations, while residual quantization converts them into multiple stages of discrete semantic codes that are transmitted concurrently over the same frequency band. Experimental validation in a 5\,m, 450\,nm laser based \gls{uowc} system demonstrated that the proposed scheme significantly outperformed the conventional \gls{jpeg} and \gls{ldpc} baseline, achieving up to a 4.68\,dB improvement in \gls{psnr} at comparable frame rates and a 329\% increase in frame rate at comparable reconstruction quality. The system attained a transmission rate of 4,394,531 image instances per second at a resolution of \(64 \times 64\).

Another approach to providing semantic information is to weight the semantic encoding features based on the channel state. Hu \textit{et al.}~\cite{11005395} proposed a semantic communication system for wireless image transmission that adapts to varying channel conditions. Their framework, termed ConvSC, uses ConvNeXt as the backbone within a joint source channel coding architecture and directly transmits analog semantic symbols without digital quantization. A semantic adaptive module combines channel and spatial attention to adjust the weights of semantic features according to the channel state information. The model was evaluated over additive white Gaussian noise, Rayleigh fading, and \gls{uowc} channels. On the UFO-120 dataset over the \gls{uowc} channel, the proposed framework achieved approximately 25.8~dB \gls{psnr} and 8.7~dB MS-\gls{ssim} at the highest evaluated \gls{snr}, with an average channel bandwidth ratio of 1/16. The model contains 20.0 million parameters, requires 113~GFLOPs, and has an inference time of 73~ms on an RTX 3090 GPU.

To overcome the bandwidth limitation of \gls{uowsc} systems, Xu \textit{et al.}~\cite{11389779} introduced \gls{ofdm} modulation into semantic communication for \gls{uowc} for the first time. The proposed system employs Swin Transformer blocks for semantic feature extraction and a \gls{spl} scheme comprising an entropy model driven symbol reordering strategy and a lightweight power allocation network. The \gls{spl} scheme adaptively allocates subcarriers and power according to semantic importance and channel conditions. Experimental validation on a 1\,m \gls{uowc} platform demonstrated superior \gls{psnr} and \gls{ssim} performance compared with conventional \gls{ofdm} based \gls{uowc} and CAP based \gls{uowsc} schemes. At a transmission bandwidth of 50~MHz, the \gls{spl} module improved semantic spectrum efficiency by up to 23\% and achieved a CR saving of 18.59\% compared with \gls{uowsc} without \gls{spl}. Across bandwidths from 10 to 50~MHz, it achieved CR savings of 66.21\% to 71.50\% compared with BPL based \gls{uowc}, while increasing the computational complexity by only 1.20~GFLOPs, or approximately 3.2\%.

\begin{table}[tp]
\caption{Performance Comparison of Semantic Communication Systems for Underwater Optical Wireless Communications}
\label{tab:semantic_performance}
\centering
\footnotesize
\setlength{\tabcolsep}{3pt}
\renewcommand{\arraystretch}{1.15}
\begin{tabular}{@{}
>{\raggedright\arraybackslash}p{1.4cm}
>{\raggedright\arraybackslash}p{1.4cm}
>{\raggedright\arraybackslash}p{1.15cm}
>{\raggedright\arraybackslash}p{0.55cm}
>{\raggedright\arraybackslash}p{1.55cm}
>{\raggedright\arraybackslash}p{1.55cm}@{}}
\toprule
\textbf{Reference} &
\textbf{Model} &
\textbf{Dataset} &
\textbf{CR} &
\textbf{Reported PSNR} &
\textbf{Reported SSIM} \\
\midrule

\cite{Xu:24}
& ResNet + LSTM
& EUVP
& 1/16
& 26.92\% improvement
& 51.90\% improvement \\

\cite{10729883}
& Swin Transformer
& EUVP, UIEB, RUIE
& --
& 2.09~dB gain (BD-PSNR)
& 0.0864 gain (BD-SSIM) \\

\cite{11389779}
& Swin Transformer
& EUVP, UIEB, RUIE
& 1/6
& 25.462~dB
& 0.9417 \\

\cite{11288833}
& VAE
& EUVP, UIEB, LSUI
& 1/48
& 22.57~dB
& 0.609 \\

\cite{11005395}
& ConvSC 
& UFO-120
& 1/16
& $\sim$25.8~dB
& $\sim$8.7~dB MS-SSIM \\

\cite{Lin:26}
& \acrshort{rq}-\acrshort{vae}
& EUVP
& --
& 28.89~dB
& -- \\
\bottomrule
\multicolumn{6}{l}{\scriptsize CR: Compression Ratio}\\
\end{tabular}
\vspace{3pt}
\end{table}

Table~\ref{tab:semantic_performance} summarizes the key performance metrics across the surveyed semantic communication systems. Several trends emerge from the comparison. First, the proposed systems generally outperform their respective conventional separation based baselines under the evaluated channel and operating conditions, demonstrating the potential of joint source channel semantic coding for \gls{uowc}. Second, Transformer based architectures~\cite{10729883,11389779} provide flexible semantic compression and adaptation by incorporating channel state information, adaptive code rates, and subcarrier resource allocation. Third, the \acrshort{rq}-\gls{vae} approach~\cite{Lin:26} reported a \gls{psnr} of 28.89~dB and the corresponding image transmission rate under its evaluated dataset, compression setting, and channel conditions. Fourth, the \gls{vae}-based system~\cite{11288833} employs a fixed 48-fold latent compression and maintains image reconstruction beyond the operating range of the conventional baseline, although its reconstruction quality is lower than that of some more recent approaches. However, direct performance ranking remains difficult because the studies use different datasets, channel models, compression ratios, transmission distances, and hardware platforms. This highlights the need for a unified benchmark for semantic \gls{uowc} systems.

\section{Receiver Design and Signal Recovery}
\label{sec:receiver}
Receiver design in \gls{uowc} systems faces 
compounding challenges from turbulence-induced 
fading, hardware nonlinearities in \glspl{spad} 
and photodetectors, and severe \gls{isi} at 
high data rates. \gls{ml} methods address these 
challenges by learning signal representations 
directly from data, bypassing the need for 
explicit channel models.

\subsection{Signal Detection and Demodulation}
\label{subsec:receiver_detection}
Deep learning addresses challenges in the receiver component, ranging from channel impairments to hardware nonlinearities. This subsection reviews representative approaches organized into two thematic areas: simulation-based joint detection frameworks and experimental demodulation methods.

\subsubsection{Detection Frameworks}
\label{subsubsec:receiver_joint}

A joint framework was proposed in \cite{11555415} for
\gls{uvlc}, using a \gls{lstm} network to jointly perform channel estimation and signal detection. The \gls{lstm} treats the underwater optical channel as a black box and learns the mapping from received noisy signals to transmitted symbols through offline training. The network consists of two \gls{lstm} layers, a fully connected layer, and a dropout layer for regularization. Training data are generated
for clear, coastal, and turbid water types at link distances of
1, 3, 10, and 30~m and across \gls{snr} levels ranging from
20 to 70~dB. Simulation results show that the \gls{lstm}-based
detector achieves \gls{ber} performance close to that of optimal maximum likelihood detection with perfect \gls{csi}, demonstrating the ability of the \gls{lstm} network to exploit temporal dependencies associated with absorption and scattering for reliable signal detection.

While the above frameworks address channel-induced impairments, hardware nonlinearities at the receiver introduce an additional layer of complexity. Jiang \textit{et al.}~\cite{8962099} addressed this challenge in \gls{spad}-based \gls{uowc} by designing a two-connected \gls{mlp} architecture integrated into the receiver, where the first subnetwork compensates for channel distortions and the second performs demodulation. The network is trained using data generated by a numerical underwater optical channel model and non-Poisson statistical models for \gls{spad} dead time effects. Simulation results demonstrate significant \gls{ber} improvement over both conventional \gls{llr}-based demodulation and a single \gls{mlp} combined with \gls{llr} demodulation.

\subsubsection{Demodulation Methods}
\label{subsubsec:receiver_experimental}

\begin{figure}[t]
    \centering
    % Source: figures/uwoc-dbn-demodulator.svg; regenerate the PDF with
    %   rsvg-convert -f pdf -o figures/uwoc-dbn-demodulator.pdf figures/uwoc-dbn-demodulator.svg
    \includegraphics[width=\columnwidth]{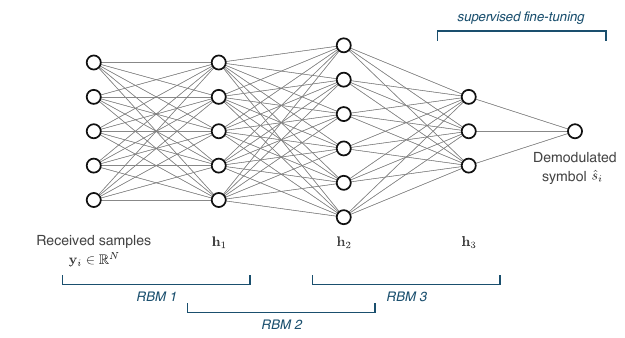}
    \caption{Architecture of the DBN demodulator. Three stacked RBMs perform layer-wise unsupervised feature extraction, followed by a finetuning layer with supervised backpropagation.}
    \label{fig:dbn}
\end{figure}

The authors in \cite{9209913} introduced an experimental investigation of \gls{dbn}-based demodulation using a practical \gls{uowc} platform and a measured dataset containing ten modulation schemes. This earlier work formed the basis for the broader experimental study later presented in \cite{10495815}. The proposed \gls{dbn} demodulator consists of three sequentially stacked \glspl{rbm}, which perform unsupervised layerwise feature learning, followed by supervised fine-tuning through backpropagation, as illustrated in Fig.~\ref{fig:dbn}. Experimental results show that the \gls{dbn} demodulator achieves the highest demodulation accuracy among the evaluated methods for 16-\gls{qam} signals. For DCO-\gls{ofdm} signals, it outperforms maximum likelihood classification, with both methods reaching 100\% accuracy when the received optical power is at least \(-16\) dBm. For 16-\gls{qam}, all evaluated methods approach 100\% accuracy when the received optical power is at least \(-19\) dBm.

Building on this earlier work, the authors in \cite{10495815} extended the experimental evaluation using the same 7-m water tank platform, which employs a blue laser diode transmitter and an \gls{apd} receiver. They publicly released the measured dataset comprising ten modulation schemes and evaluated two \gls{ml}-based demodulators. The first is a \gls{dbn}-based demodulator with three hidden layers, in which stacked \glspl{rbm} are trained through unsupervised pretraining and subsequently fine-tuned using supervised backpropagation. The second is an AdaBoost-based demodulator that combines multiple \gls{knn} weak classifiers through adaptive sample weighting and weighted voting. The proposed demodulators are compared with benchmark methods, including \gls{svm}, maximum likelihood classification, \gls{cnn}, and naive Bayes classification. Experimental results show that the AdaBoost-based demodulator generally achieves the highest demodulation accuracy, demonstrating the potential of ensemble learning for practical \gls{uowc} signal demodulation. The released dataset also provides a useful benchmark for future data-driven demodulation studies.

Another \gls{dbn} architecture was proposed for signal reconstruction and denoising in \gls{uowc} receivers. Yousef and El-Eraki~\cite{Yousef2026DBN} presented a \gls{dbn}-based receiver architecture that integrates a \gls{mc} channel simulator with a signal-to-image pixelization algorithm. The algorithm transforms received one-dimensional time series signals into two-dimensional grayscale feature maps, allowing the \gls{dbn} to learn spatial representations of scattering tails and noise patterns. The \gls{dbn} is pretrained using contrastive divergence and subsequently fine-tuned through supervised backpropagation. The framework considers four water types using \gls{mc}-generated channel realizations. The reported analysis projects a coding gain of approximately 7.8--12~dB over conventional maximum likelihood estimation detection under turbid harbor water conditions. However, these gains are based on simulated and projected performance rather than independent experimental validation.

These studies indicate that \gls{dbn}-based receivers can learn useful signal representations through unsupervised pretraining followed by supervised fine tuning. Shuai \textit{et al.}~\cite{10495815} and Ma \textit{et al.}~\cite{9209913} demonstrated this capability using measured \gls{uowc} data, while Yousef and El-Eraki~\cite{Yousef2026DBN} explored its potential for signal reconstruction and denoising using simulated channel data.

\gls{ml} classifiers have also been investigated as demodulators. Nennouche \textit{et al.}~\cite{10636416} proposed a \gls{knn} demodulator for \gls{uowc} using the experimental dataset introduced by Ma \textit{et al.}~\cite{9209913}. The study evaluated six single-carrier modulation schemes across a range of received optical power levels and also considered DCO \gls{ofdm}. The \gls{knn} demodulator achieved 100\% classification accuracy for \gls{ook}, 4 \gls{ppm}, and 4 \gls{qam} at sufficiently high received optical power when 64 samples per symbol were used. For 16 \gls{qam} and DCO \gls{ofdm}, its performance was comparable to that of the \gls{dbn} demodulator in \cite{9209913}, while requiring lower computational complexity. These results show that classical \gls{ml} algorithms remain competitive baselines for practical \gls{uowc} signal detection.

Beyond conventional modulation schemes, \gls{cnn} demodulators have also been applied to \gls{oamsk} signals. Cui \textit{et al.}~\cite{Cui2019OAM} experimentally demonstrated a seven-layer \gls{cnn} decoder for 16-ary \gls{oamsk} in a 1-m underwater tank. The decoder classifies the intensity patterns produced by superposed \gls{oam} modes. In clean salty water, the reported decoding accuracy ranged from 98.62\% to 99.25\%, depending on the selected mode set. In turbid salty water, accuracies above 99\% were achieved for several mode sets when sufficiently high input image resolution was used. Under oceanic turbulence, the decoding accuracy decreased markedly as the turbulence strength increased, particularly for larger \gls{oam} modes. For the strongest tested turbulence condition, the 16-ary system achieved approximately 60.78\% to 71.03\% accuracy, depending on the input image resolution. These results highlight the sensitivity of high-order \gls{oamsk} decoding to severe turbulence and show that increasing the input resolution or reducing the modulation order can improve robustness.

A generalization framework across the four canonical water types was presented in~\cite{9817578}, where the authors applied the \gls{maml} algorithm to a \gls{uowc} receiver comprising two cascaded \gls{mlp} networks. The first subnetwork compensates for channel distortion, while the second performs signal demodulation. The framework was evaluated using \gls{mc} simulated channels under matched and mismatched water type conditions. In the mismatched setting, the model was trained using three water types and adapted to a fourth unseen water type. The evaluated system employed 64 subcarrier DCO \gls{ofdm} transmission at 1~Gbps with 4 \gls{qam} and 16 \gls{qam} modulation. Simulation results showed that the MAML based receiver achieved lower \gls{ber} than conventional \gls{ls} \gls{ofdm}, complex \gls{dnn}, and pruned \gls{dnn} receivers. It also exhibited faster adaptation and greater robustness under mismatched channel conditions, where the performance of the benchmark receivers degraded. The proposed receiver extended the supported transmission distance, particularly in turbid harbor water, and retained its performance advantage when 16 \gls{qam} was employed.

The detection and demodulation methods reviewed in this subsection demonstrate that learning methods can address diverse receiver challenges in \gls{uowc} systems. The joint framework in~\cite{11555415} achieved \gls{ber} performance close to optimal maximum likelihood detection. Classical \gls{ml} demodulators also provided competitive performance. The AdaBoost demodulator in~\cite{10495815} generally achieved the highest demodulation accuracy among the evaluated methods, whereas the demodulator using \gls{knn} in~\cite{10636416} achieved performance comparable to the \gls{dbn} receiver with lower computational complexity. The receiver using a \gls{dbn} in~\cite{9209913} achieved high accuracy using measured experimental data, while Yousef and El-Eraki~\cite{Yousef2026DBN} investigated \gls{dbn} signal reconstruction and denoising using simulated channels. The receiver using \gls{maml} in~\cite{9817578} demonstrated adaptation across different water types. Generalization across modulation formats, continuously varying channel conditions, and different hardware platforms remains insufficiently reviewed, motivating the learning assisted equalization methods discussed next.

\subsection{Learning-Assisted Equalization}
\label{subsec:receiver_equalization}

Channel distortions and hardware nonlinearities 
in \gls{uowc} systems degrade received signal 
quality in ways that conventional linear 
equalizers cannot fully address. Neural 
networks have been proposed to 
learn the nonlinear mapping between distorted 
and original signals directly from data.

Wang \textit{et al.}~\cite{11109827} proposed a neural network architecture that jointly performs equalization and decoding. The architecture processes $2\times$ oversampled \gls{ook} waveforms by learning inter-sample correlations between adjacent symbols. The receiver uses four samples as input: two samples from the current symbol, one from the preceding symbol, and one from the following symbol, thereby capturing transition information across symbol boundaries. The \gls{dnn} employs a shallow architecture comprising two hidden layers with 64 and 32 neurons, respectively, and an eight-neuron output layer for classifying three-symbol sequences. The network is trained using simulated signals and subsequently evaluated using experimentally measured signals to reduce overfitting to device-specific noise and improve its generalization to practical underwater channels. Experimental results show that the proposed receiver achieves received optical power gains of 2.3~dB and 1.1~dB over conventional threshold detection and maximum likelihood estimation, respectively, without introducing redundant bits.

Another \gls{nn} architecture was proposed by Du \textit{et al.}~\cite{Du2024SemiTNN}, who developed a \gls{semitnn} post-equalizer for \gls{pam}4 \gls{uowc} systems to address the limited availability of labeled data in dynamic underwater environments. The \gls{semitnn} employs a dual-drop strategy that combines neuron drop and threshold drop with interleaved consistency regularization, allowing effective training with only 5--10\% labeled data. Experimental results over a 14-m, 4-Gbps underwater link showed that the \gls{semitnn} using 5\% labeled data achieved nearly the same \gls{ber} performance as fully supervised training. When the transmission distance was extended to 56~m, the \gls{semitnn} using 10\% labeled data achieved a data rate of 4.26~Gbps below the hard-decision \gls{fec} threshold, exceeding the rates achieved by Volterra nonlinear and linear feedforward equalizers by 90~Mbps and 150~Mbps, respectively. The authors reported this as the first implementation of a \gls{semitnn} combined with interleaved consistency regularization for high-speed \gls{uowc}, demonstrating the potential of semi-supervised learning to reduce labeled training overhead in practical underwater equalization systems.

For high-order modulation schemes, hardware-induced nonlinearities present a more severe equalization challenge. Cai \textit{et al.}~\cite{9880586} proposed a post-equalization method using a \gls{bgru} network together with the 64-\gls{apsk} modulation format for \gls{uvlc}. Unlike the conventional equalizers, which can compensate for limited nonlinear effects but degrades under more complex nonlinear distortion, the \gls{bgru} exploits bidirectional recurrent processing to capture dependencies from both preceding and subsequent samples. A sliding-window input strategy provides each target symbol with contextual information from neighboring samples, thereby mitigating inter-symbol interference. The method was experimentally validated over a 1.2-m water tank using a blue \gls{led} transmitter and a \gls{pin} photodetector receiver. Experimental results showed that the \gls{bgru} outperformed conventional equalizers as the drive voltage increased and the nonlinear distortion became more severe, achieving a maximum Q-factor improvement of 2.4~dB at \(V_{\mathrm{PP}}=0.9\)~V, a bias current of \(150\)~mA, and a bit rate of 3~Gbps. The combined 64-\gls{apsk} and \gls{bgru} scheme also achieved a maximum transmission rate of 3.19~Gbps over the 1.2-m link.

Complementing recurrent approaches, Wang \textit{et al.}~\cite{10089049} investigated \gls{deepesn} as a signal equalizer for high-speed \gls{uowc} systems supporting both \gls{pam} and \gls{qam}-\gls{ofdm} modulations. The \gls{deepesn} processes temporal signal states through multiple recurrent reservoir layers, while only the output weight matrix is trained using \gls{svd}, providing a computationally efficient training procedure while retaining strong temporal modeling capability. The system was experimentally validated over a 40.5~m underwater optical link using a 520~nm green laser and an \gls{apd} receiver. For \gls{pam}4 modulation, the \gls{deepesn} exhibited an advantageous behavior in which higher-data-rate signals benefited more from the equalization process, reaching the adopted \gls{ber} threshold of \(10^{-3}\) at a lower received optical power for 166.67~Mbps than for 100~Mbps. For \gls{qam}-\gls{ofdm}, the \gls{deepesn} consistently outperformed recursive \gls{ls} equalization across data rates from 100 to 130~Mbps, reaching the \(10^{-3}\) \gls{ber} threshold at received optical powers of \(-20.2\), \(-19.6\), \(-18.1\), and \(-17\)~dBm for 100, 110, 120, and 130~Mbps, respectively.

Building on channel-estimation and learning-based \gls{ber} prediction approaches, Salama \textit{et al.}~\cite{Salama2025LowSNR} evaluated several \gls{dl} architectures combining \gls{cnn}, \gls{tcn}, \gls{rnn}, \gls{lstm}, and attention mechanisms using simulated \gls{uowc} data. Among the evaluated architectures, the TCN-LSTM-AM model achieved the best overall prediction performance, attaining an accuracy of 97.99\% and reducing \gls{mse}, \gls{rmse}, and \gls{mae} by 66.15\%, 41.90\%, and 40.96\%, respectively, relative to the baseline \gls{cnn} model. For \gls{qpsk}-\gls{ofdm} combined with deep echo state network (DESN) processing, the source reports \gls{ber}-performance improvements of 29.4\%, 41.17\%, 53.3\%, and 71.4\% at transmission ranges of 100, 110, 120, and 130~m, respectively, compared with the reference results used in that study. These simulation results indicate that combining temporal convolution, recurrent modeling, and attention can improve \gls{ber} prediction and reduce prediction errors under the evaluated low-\gls{snr} conditions.

\gls{cnn} decoders have also been applied to \gls{oam} multiplexing in \gls{uowc}. Huang \textit{et al.}~\cite{Huang2026OAM} proposed and numerically evaluated a \gls{cnn} decoding method that accounts for the combined effects of absorption, scattering, turbulence, noise, and diffraction on multiplexed \gls{oam} beams. The \gls{cnn} is trained to recover the transmitted information directly from intensity maps of distorted multiplexed \gls{oam} light without requiring explicit channel estimation or equalization. The proposed method generally outperformed the traditional Gerchberg--Saxton algorithm, with the largest advantages observed under moderate and strong turbulence, while also requiring lower computational complexity.

\begin{table*}[t]
\caption{Neural Network Equalizers for \gls{uowc}}
\label{tab:uwoc_equalizer_synthesis}
\centering
\footnotesize
\setlength{\tabcolsep}{4pt}
\renewcommand{\arraystretch}{1.15}

\begin{tabularx}{\textwidth}{
@{}>{\raggedright\arraybackslash}p{1.6cm}
>{\raggedright\arraybackslash}p{2.1cm}
>{\raggedright\arraybackslash}p{2.2cm}
>{\raggedright\arraybackslash}X
>{\raggedright\arraybackslash}p{2.5cm}@{}
}
\toprule
\textbf{Reference} &
\textbf{Method} &
\textbf{Signal Format} &
\textbf{Main Result} &
\textbf{Evaluation} \\
\midrule

\cite{11109827}
&
Shallow \gls{dnn}
&
$2\times$ oversampled \gls{ook}
&
Joint equalization and decoding; 2.3~dB and 1.1~dB received-power gains over threshold detection and maximum likelihood estimation.
&
Experimental
\\
\addlinespace

\cite{9880586}
&
\gls{bgru}
&
64-\gls{apsk}
&
Maximum Q-factor improvement of 2.4~dB and a transmission rate of 3.19~Gbps.
&
1.2-m experimental link
\\
\addlinespace

\cite{10089049}
&
\gls{deepesn}
&
\gls{pam}4 and \gls{qam}-\gls{ofdm}
&
Outperformed recursive \gls{ls} equalization and trained only the output weights using \gls{svd}.
&
40.5-m experimental link
\\
\addlinespace

\cite{Salama2025LowSNR}
&
TCN-LSTM-AM
&
\gls{pam} and \gls{qpsk}-\gls{ofdm}
&
Achieved 97.99\% accuracy; the source reports a
\gls{ber}-performance improvement of up to 71.4\%
at 130~m relative to its reference results. &
Simulation
\\
\addlinespace

\cite{Huang2026OAM}
&
\gls{cnn}
&
Multiplexed \gls{oam}
&
Directly decoded intensity maps and outperformed the Gerchberg--Saxton algorithm under moderate and strong turbulence.
&
Numerical simulation
\\
\addlinespace

\cite{Du2024SemiTNN}
&
\gls{semitnn}
&
\gls{pam}4
&
Reached 4.26~Gbps using only 5--10\% labeled data.
&
14-m and 56-m experiments
\\
\bottomrule
\end{tabularx}
\end{table*}
In summary, neural network equalizers have emerged as effective solutions for mitigating nonlinear distortions and channel impairments in \gls{uowc} systems. As summarized in Table~\ref{tab:uwoc_equalizer_synthesis}, Wang \textit{et al.}~\cite{11109827} demonstrated that a shallow \gls{dnn} can jointly perform equalization and decoding of \(2\times\) oversampled \gls{ook} signals by learning inter-sample correlations across adjacent symbols. Recurrent architectures such as the \gls{bgru} in~\cite{9880586} effectively capture temporal dependencies and outperform \gls{lms} and Volterra equalizers under severe nonlinear distortion. Deep Echo State Networks~\cite{10089049} provide a computationally efficient alternative because only the output weight matrix is trained. Hybrid architectures combining \gls{tcn}, \gls{lstm}, and
attention mechanisms~\cite{Salama2025LowSNR} also improved \gls{ber} prediction under the simulated transmission ranges, with the source reporting improvements of up to 71.4\% relative to its reference results. \gls{cnn} decoders have also been applied to \gls{oam} multiplexing~\cite{Huang2026OAM}, enabling direct recovery of transmitted information from distorted intensity maps under underwater turbulence. Finally, Du \textit{et al.}~\cite{Du2024SemiTNN} proposed a semi-supervised architecture that enforces prediction consistency across different dropout configurations, thereby reducing the required amount of labeled training data. Nevertheless, most of the reviewed methods rely on conventional feedforward and recurrent neural architectures. Other model families, particularly Transformers, remain comparatively underexplored for \gls{uowc} equalization and could offer improved long-range dependency modeling and adaptability across varying channel conditions.

\section{Link Alignment and Adaptive Control}
\label{sec:link_alignment}
Many machine learning methods have been explored for link alignment in \gls{uowc} systems. \gls{cnn} classifiers have been widely used for misalignment estimation, reinforcement learning has been applied for beam optimization, and various algorithms have been developed for \gls{apt}. These approaches collectively address the critical challenge of maintaining link reliability in dynamic underwater environments.

\subsection{Misalignment Estimation}

\gls{dl} has been applied to mitigate link misalignment without requiring \gls{csi}. Lu \textit{et al.}~\cite{9455389} proposed a misalignment-robust blind receiver using \glspl{cnn} for \gls{mimo} \gls{uowc} systems. The receiver comprises a \gls{cnn} combiner that determines the weights of signals received by multiple photodetectors and a \gls{cnn} demodulator that recovers the transmitted \gls{ook} symbols. Trained using zenith angles from $0^\circ$ to $10^\circ$, the CNN-MBR can handle link misalignment without prior knowledge of its level and generalize to untrained angles within the training range. Simulation and experimental results for coastal and harbor water channels showed performance close to ideal maximum-ratio combining and better performance than maximum-ratio combining using practical minimum mean-square-error channel estimation.

Link alignment is challenged by variations in light-source appearance under different misalignment conditions. Jia \textit{et al.}~\cite{10663260} proposed a cascaded \gls{dl} detector comprising a YOLOv8s object detector that identifies the light-spot region and an RTMPose-t keypoint detector that localizes the light-source coordinates. The method supports terminal-camera distances from 1~m to 32~m, terminal deviation angles from $0^\circ$ to nearly $180^\circ$, and camera deviation angles exceeding $45^\circ$, while remaining robust to surface reflections, motion blur, background-light interference, and dim light spots. With 8.1~GFLOPs and an inference time of 9.1~ms on an NVIDIA RTX 3060 \gls{gpu}, the method has potential for real-time deployment. The accompanying ULDB dataset contains 2,200 images covering diverse misalignment scenarios. Compared with the grayscale centroid method, the proposed detector reduced the average positioning error from 142.08 pixels to 4.66 pixels.

For spatially multiplexed twin-beam \gls{uowc} systems, Tanaka \textit{et al.}~\cite{10926134} developed a geometric alignment algorithm using a \gls{cnn} to simultaneously detect and compensate for axis and angular misalignment from a single camera image. The algorithm applies sequential coarse and fine tuning, extracting aberration and positional features from the beam cross-sectional images through \gls{cnn} classification. Experiments over a 2.4-m underwater channel using twin 520-nm flat beams and time-domain hybrid \gls{pam} signals showed that the algorithm reduced the average \gls{ber} from $4.4\times10^{-1}$ to $2.8\times10^{-3}$, below the 7\% \gls{hdfec} limit of $3.8\times10^{-3}$.

The \gls{cnn} architectures proposed for misalignment mitigation in \gls{mimo} and spatially multiplexed twin-beam \gls{uowc} systems~\cite{9455389,10926134} were validated using measured or experimental data under diverse underwater conditions. The cascaded detector combining YOLOv8s and RTMPose-t~\cite{10663260} further extended light-source localization capabilities, supporting a wide range of misalignment conditions with real-time inference. Collectively, these studies demonstrate the potential of \gls{dl} to overcome the limited robustness of conventional image-processing and model-based alignment techniques under challenging underwater conditions.

\subsection{Beam Optimization}

\gls{rl} has been widely adopted for adaptive beam control in \gls{uowc} systems, addressing the dynamic misalignment caused by platform motion, sea waves, and \gls{auv} mobility. These methods can be broadly categorized by their control objectives: divergence adaptation, full beam steering, power optimization, and multi-agent coordination.

Early works focused on single-objective beam control. Romdhane and Kaddoum~\cite{9770197} compared Q-learning and \gls{sarsa} for beamwidth adaptation, beam-orientation adaptation, and their joint optimization. \gls{sarsa} generally converged faster than Q-learning, achieving a success rate of approximately 93\% for beamwidth adaptation after about 100 iterations. However, for beam orientation adaptation, Q-learning achieved a higher final success rate, while \gls{sarsa} converged more rapidly. All proposed \gls{rl} methods outperformed the static uncertainty-disk approach across four water types, with the \gls{snr} ranging from 32--44~dB in pure seawater to 4--17~dB in turbid harbor water.

To address time-varying misalignment caused by sea-wave-induced vessel motion, Ishida \textit{et al.}~\cite{11003393} combined an \gls{lstm} model for wave prediction with a \gls{dqn} for real-time beam control in \gls{auv}--\gls{asv} links. Two strategies were evaluated: beam-divergence adaptation, which converged after approximately 20 iterations, and joint beam-divergence and steering control, which required approximately 50 iterations. The \gls{lstm}--\gls{dqn} framework achieved higher and more stable average rewards than \gls{dqn}-only and heuristic methods. In scenarios with a moving \gls{auv}, variations in link distance increased reward fluctuations and made the learning process less stable.

Building on single-agent approaches, subsequent works introduced multi-agent frameworks for joint beam optimization. Shin \textit{et al.}~\cite{10215370} proposed a two-phase two-agent \gls{drl} algorithm in which an inner agent adjusts the beam-divergence angle based on the instantaneous \gls{snr}, while an outer agent selects the transmission-power level based on the long-term \gls{snr}. The joint approach reduced power consumption by approximately 29\% while satisfying a required \gls{snr} of 16~dB. A two-step two-agent \gls{drl} framework~\cite{10537612} sequentially optimized beam orientation and beam divergence, achieving higher \gls{snr} and more stable link alignment than the considered benchmark methods using measured \gls{usv} movement data. More recently, a three-agent framework~\cite{11257798} jointly optimized beam orientation, beam divergence, and transmission power for communication between fixed seabed sensors and a mobile \gls{usv}, improving link-maintenance probability and energy efficiency.
For mobile multi-\gls{auv} scenarios, Li \textit{et al.}~\cite{10753441} proposed a \gls{drl} framework combining \gls{ddpg} with \gls{ekf} for cooperative movement and TD3 for adaptive transmission-power control. The \gls{ddpg} strategy achieved more stable light intensity and relative positioning than leader–follower and heuristic alignment methods. After adaptive power adjustment, the end-to-end \gls{ber} decreased from the order of \(10^{-3}\) to \(10^{-7}\). The authors identified this as the first study focusing on mobile optical communication using multiple \glspl{auv}.

These \gls{drl} methods demonstrate that learning-based control can effectively maintain \gls{uowc} links under dynamic conditions. However, most studies have been evaluated primarily in simulation settings, and real world deployment on \glspl{auv} and \glspl{usv} under turbulent water conditions and practical hardware constraints remains an open challenge.

\subsection{Acquisition, Pointing, and Tracking (\gls{apt})}

For \gls{apt} in \gls{uowc}, Kong \textit{et al.}~\cite{kong2024deep} proposed an \gls{apt} system using YOLOv5s and a 12\,cm $\times$ 12\,cm solar-panel receiver with a large detection area. The system operates in three stages: acquisition, which locates the target direction within an uncertain area; pointing, which rotates the transceiver until the target is centered in the image; and tracking, which continuously maintains pointing to preserve the communication link. YOLOv5s was trained using 4,500 images, including 3,002 underwater and 1,498 indoor images. Experiments in a 3-m water tank demonstrated average pointing and tracking times of 5.97~ms and 229~ms, respectively, under varying turbidity, bubble-induced turbulence, and background-light conditions. Preliminary mobile experiments demonstrated target tracking at speeds of up to 5~cm/s, while the \gls{ber} remained below the \gls{fec} limit for a lateral offset of 5~cm at all tested speeds.

\gls{rl} has been employed with acoustic navigation for optical beam alignment between \glspl{auv}. Weng \textit{et al.}~\cite{weng2022reinforcement} formulated the alignment problem as a partially observable Markov decision process in which the transmitting \gls{auv} must maintain a specific relative position and orientation to establish an optical link. Acoustic navigation provides relative position and orientation information without requiring beam directors, light intensity sensors, or scanning algorithms. The state space includes relative position, orientation, and velocity, while the actions control surge velocity, yaw angular velocity, acoustic ranging requests, and optical transmitter activation. A \gls{sac} algorithm using curriculum learning and reward shaping learns the alignment policy, while a particle filter estimates the system state from acoustic observations. The learned policy achieved a 97.53\% probability of maintaining the \gls{los} link for ten consecutive time steps in simulation and was successfully transferred to the real \gls{auv} Tri-TON 2.

Another tracking controller using \gls{sac} was proposed in~\cite{WENG2025121047} to enable an \gls{auv} to track a seafloor platform and establish a short-range, three-dimensional \gls{uowc} link. The controller adjusts the surge velocity, yaw angular velocity, commanded depth, and elevation of the scanning device to reduce pointing errors and energy consumption. Trained in simulation and deployed on the \gls{auv} Tri-TON in sea experiments, the system completed link pointing in 18~s while maintaining the distance error within 1.10~m, the bearing angle error within 10.54$^\circ$, and the elevation angle error within 1.70$^\circ$.

These \gls{apt} approaches demonstrate that both \gls{rl} and vision-based methods can effectively address the challenges of target acquisition, beam alignment, and tracking in \gls{uowc} systems. Notably, the framework proposed by Weng \textit{et al.}~\cite{weng2022reinforcement} stands out as one of the few studies that successfully transferred a learned policy from simulation to a real \gls{auv} (Tri-TON 2), a feat also demonstrated in their subsequent work on seafloor-platform tracking~\cite{WENG2025121047}.

\section{Beyond Communication: Emerging Applications}
\label{sec:emerging_applications}
This section explores the application of machine learning 
techniques in \gls{uowc} beyond conventional 
communication systems, including optical wireless power 
transfer, underwater optical sensing and imaging, localization and positioning, and intelligent underwater sensor networks.

\subsection{Underwater Wireless Power Transfer}

Optical signals carry not only information but also energy, 
enabling simultaneous wireless power transfer and communication for underwater devices including \glspl{auv}, \glspl{rov}, and \gls{iout} nodes.
Khalfet \textit{et al.}~\cite{11493549} proposed the \gls{ciecl} framework, a cooperative \gls{gan}-inspired architecture that learns capacity-achieving input distributions for \gls{slipt} systems over lognormal fading channels under peak-power, average-power, and nonlinear energy-harvesting constraints. The generator approximates the capacity-achieving input distribution, while the discriminator estimates mutual information by distinguishing paired from unpaired channel input--output samples. Penalty terms enforce the \gls{slipt} constraints. By avoiding the discretization required by the conventional Blahut--Arimoto algorithm, the framework directly handles nonlinear energy-harvesting constraints and achieves higher mutual information under the considered conditions.
Building on this theoretical foundation, Shin \textit{et al.}~\cite{shin2024adaptive} developed a hierarchical \gls{drl} algorithm for adaptive \gls{slipt} control in underwater optical wireless communication. In their system, an \gls{rov} uses an LD-based downlink to simultaneously deliver control data and optical power to an underwater sensor, which subsequently transmits sensing data to the \gls{rov} through an \gls{led}-based uplink. The proposed hierarchical \gls{dqn}--\gls{ddpg} framework employs two cooperating \gls{rl} agents: a \gls{dqn} agent at the \gls{rov} that determines the beam-divergence angle to maximize the received optical power while maintaining link continuity, and a \gls{ddpg} agent at the sensor that optimizes the time-switching and power-splitting ratios to maximize energy harvesting while satisfying the required spectral efficiency. Extensive simulations demonstrated at least an 11\% improvement in energy-harvesting performance over benchmark \gls{slipt} algorithms.

Beyond optical \gls{slipt}, Chen \textit{et al.}~\cite{11185120} proposed a hybrid mechanism for jointly identifying mutual inductance and load in LCC–S--compensated \gls{uwpt} systems. A multifrequency excitation strategy breaks the linear dependence and parameter coupling inherent in the compensation topology, enabling accurate identification using only two transmitter-side voltage features. An \gls{ann} is pretrained using mechanism-derived data to construct a source-domain model, which is then fine-tuned through transfer learning using limited experimental data, thereby improving generalization across different operating conditions. Experimental validation on a 1\,kW/85\,kHz platform in air and three saltwater conditions demonstrated prediction errors below 3\% and identification times within 2.72~ms.

Together, these studies advance underwater power transfer from complementary directions. However, adaptive \gls{slipt} control under dynamic underwater conditions, including turbulence, platform motion, and variations in water type, remains an open challenge.

\subsection{Underwater Optical Sensing and Imaging}

\gls{dl} has transformed underwater optical
sensing and imaging for high quality image restoration,
enhancement, and object detection in turbid environments where conventional methods suffer from light scattering and beam attenuation. This section reviews recent advances organized by technical approach.
\subsubsection{Underwater Image Enhancement}

Recent advances in underwater image enhancement have evolved beyond traditional image processing techniques toward intelligent approaches that incorporate either physical underwater imaging models or semantic guidance. These methods aim to address challenges such as color distortion, scattering induced blur, and contrast degradation while preserving scene content, directly benefiting the visual perception capabilities of \glspl{auv} and \glspl{rov} operating in turbid environments.

Zhang \textit{et al.} proposed P\textsuperscript{2}DNet, a physics-constrained learning framework for underwater polarimetric image restoration~\cite{zhang2026p2dnet}. The architecture consists of B-Net, which estimates the global background-light intensity from the total-intensity image, and D-Net, which predicts the degree of linear polarization of backscattered light using orthogonal polarization images. To support training and evaluation, the authors introduced the NEU-PURGB dataset, containing 1,008 polarization--RGB image pairs collected from 42 scenes under five calibrated scattering levels. Experimental results demonstrated that P\textsuperscript{2}DNet achieved 26.05~dB \gls{psnr} and 0.8973 \gls{ssim}, outperforming DICAM and MIRNet-v2. Ablation studies further verified the contributions of the attention, multiscale-processing, brightness-adaptation, and physics-guided loss components.

Building on physical-modeling principles, Tan \textit{et al.}~\cite{Tan2026PGMamba} proposed PGMamba, a physical-model-guided global Mamba architecture for underwater image enhancement. The architecture integrates a \gls{sagmamba} block for long-range dependency modeling and a \gls{pmgffn} that explicitly incorporates underwater optical imaging principles. PGMamba achieved state-of-the-art performance on the UIEB and LSUI datasets. On LSUI, it achieved a \gls{psnr} of 28.66~dB and an \gls{ssim} of 0.9314, outperforming the considered \gls{cnn} and Transformer methods. This work demonstrates the potential of state-space models for physics-guided underwater image enhancement.

While physics-constrained approaches exploit physical imaging principles, Wang \textit{et al.} proposed LADAR (Language-Adaptive Dynamic Aqua Refinement), which introduces semantic guidance through natural-language prompts to improve underwater image restoration~\cite{wang2025ladar}. The framework incorporates a Dual-Stream Gated Attention module for processing positive and negative textual prompts, a Laplacian Spectral Gating module for frequency-domain feature modulation, and a Hybrid Calibration Module that integrates underwater optical priors with language-guided representations. Across multiple benchmark datasets, LADAR achieved state-of-the-art performance, including 23.95~dB \gls{psnr} and 0.915 \gls{ssim} on UIEB, 27.15~dB \gls{psnr} and 0.901 \gls{ssim} on EUVP, and the highest UCIQE score of 0.601 on UIQS. Moreover, the enhanced images improved downstream object-detection performance, increasing mAP from 70.15\% to 87.06\% using a pretrained YOLOv8 detector. 

Collectively, these approaches combine physical imaging priors, polarization information, state-space modeling, and semantic guidance to improve underwater image enhancement and restoration. However, their evaluations remain distributed across different datasets and controlled degradation conditions, making consistent comparison of their generalization across canonical water types and real-world environments challenging.

\subsubsection{Underwater Optical Detection}

Krishnan \textit{et al.}~\cite{krishnan2021optical:21} presented the first reported approach for underwater optical signal detection using multidimensional integral imaging with deep neural networks. Optical signals were temporally encoded using a 7-bit Gold code and transmitted through turbid water using a 630~nm \gls{led}, while a $3\times3$ camera array captured multiperspective video sequences that were reconstructed into three-dimensional integral-imaging videos to enhance signal visibility under scattering and partial-occlusion conditions. A \gls{cnn} combined with a bidirectional \gls{lstm} was employed to classify the transmitted symbols and idle sequences, achieving an area under the \gls{roc} curve of 0.8965 under high turbidity, substantially outperforming conventional two-dimensional imaging, which achieved 0.4805. The same architecture was adopted by Joshi \textit{et al.}~\cite{joshi2024underwater} for simultaneous object detection and optical signal classification, with YOLOv4 used for object detection. At the lowest tested turbidity level, the system achieved precisions of 0.970 and 0.932 for the shark and submarine targets, respectively. It also maintained a Matthews correlation coefficient of 0.9 for signal detection at the highest tested turbidity level, whereas conventional two-dimensional imaging yielded a coefficient of zero. These studies demonstrate the effectiveness of integral imaging in improving both optical-signal recovery and visual perception in turbid and partially occluded underwater environments.

Wang \textit{et al.}~\cite{wang2023underwater} applied YOLOv7 to underwater optical-image object detection using the URPC dataset, which contains 8,199 images spanning four marine-target categories. The YOLOv7 architecture incorporates the E-ELAN module for efficient feature extraction, the MP1 module for reducing feature loss during downsampling, and the SPPCSPC module for improving the detection of targets at different scales. Experimental results showed that YOLOv7 outperformed the considered YOLOv5 models in overall accuracy, achieving a precision of 0.847 and a recall of 0.795. The model also demonstrated robustness to target occlusion, sediment burial, color distortion, and image blur, highlighting its suitability for target-detection and recognition tasks in underwater unmanned systems.

While optical-image detectors provide high-resolution visual information, their performance degrades substantially in highly turbid environments. To address this limitation, Yu \textit{et al.} proposed AO-UOD, an acousto--optic fusion framework that combines sonar and optical imagery for robust underwater object detection~\cite{yu2025aouod}. The authors constructed a paired acousto--optic dataset comprising 8,001 image pairs across eight object categories and developed a dual-stream interactive detection network using YOLOv5s as its baseline. The proposed attention-based interactive dual-branch fusion module employs self-attention to extract texture and background information and modality attention to enable cross-modal feature interaction. AO-UOD achieved mAP@50 values of 99.91\% and 98.71\% for its optical and sonar detection heads, respectively, outperforming most of the considered unimodal methods. The framework also maintained robust detection performance under optical degradation, simulated turbidity, and sonar-speckle-noise conditions.

Overall, these studies illustrate the growing role of \gls{ml} in underwater sensing applications. Integral imaging improves optical-signal recovery and visual perception in scattering environments, modern object detectors such as YOLOv7 enable accurate visual recognition, and multimodal acousto--optic fusion further improves robustness under adverse underwater conditions in which purely optical approaches may become unreliable.

\subsection{Underwater Localization and Positioning}

Accurate localization of underwater vehicles remains challenging because of the absence of \gls{gps} signals and the limited reliability of individual sensing modalities. Recent \gls{ml} approaches have explored \gls{rl}, biologically inspired perception, and multisensor fusion to improve localization and navigation in underwater environments.

Weng and Maki proposed an \gls{auv} data-mule communication system that enables high-speed data retrieval from seafloor platforms using \gls{uowc}~\cite{10682234}. The framework combines acoustic positioning for initial localization with \gls{uowc} for short-range, high-speed data transfer. To establish the optical link, a \gls{sac} controller is employed to reduce pointing errors. The state space includes relative-position errors, orientation information, and vehicle velocities, while the action space comprises surge, sway, heave, and yaw velocity commands, together with scanning-device elevation control. Trained using \(1.5\times10^{6}\) simulated samples, the controller reduced the pointing error from 15.53~m to 1.05~m within 42.2~s and subsequently maintained it at approximately 1~m in simulation. This study demonstrates the potential of \gls{rl} for autonomous \gls{auv} maneuvering and optical-link establishment, although validation on a real platform remains future work.

Rather than relying on external positioning infrastructure, Zhang \textit{et al.} developed a bio-inspired polarization-based underwater geolocation method~\cite{zhang2025nav}. The method employs an SE-ResNet50 network to jointly process the angle of polarization, degree of polarization, and light intensity to estimate the solar zenith and azimuth angles. An iterative compression-approximation algorithm subsequently converts the estimated solar angles into geographic coordinates. In the single-location experiment, the method achieved average solar-zenith and azimuth estimation errors of \(0.169^\circ\) and \(0.147^\circ\), respectively, with more than 96\% of predictions falling within \(0.5^\circ\) of the ground truth. By exploiting naturally available underwater polarization patterns, this approach provides a promising drift-free localization method for \gls{gps}-denied underwater environments, although its robustness across locations and weather conditions remains limited.

Multi-sensor fusion has also been investigated to improve localization robustness. Tiwari \textit{et al.} proposed an autonomous underwater mine-detection framework that integrates \gls{gps} information from a surface relay, inertial measurements, side-scan sonar positioning, magnetometer measurements, and optical imagery through an \gls{ekf}~\cite{tiwari2025mine}. The \gls{ekf} provides vehicle-state estimates and spatial alignment among the sensing modalities in the presence of measurement noise and uncertainty. The resulting fused observations are subsequently processed by a modified YOLOv8-inspired detector, while the detected mine locations are supplied to a navigation controller for autonomous hazard avoidance.

These studies show the diversity of \gls{ml} methods applied to underwater localization. \gls{rl} facilitates adaptive navigation and optical link establishment, polarization approaches exploit natural environmental cues, while probabilistic multi-sensor fusion improves localization accuracy and robustness through complementary sensing modalities. Together, these techniques provide promising pathways toward reliable navigation and positioning for future autonomous underwater platforms.

\subsection{Underwater Optical Sensor Networks}

\glspl{uosn} extend the capabilities of point-to-point \gls{uowc} links to multinode architectures, enabling distributed data collection, environmental monitoring, and coordinated sensing across large oceanic regions. Key challenges in \glspl{uosn} include energy management, network-topology optimization, adaptive routing under dynamic channel conditions, and reliable data delivery in the presence of turbulence, attenuation, and link misalignment.

To address these challenges, Simon \textit{et al.}~\cite{simon2025dual} proposed a dual deep-learning framework for energy-efficient communication in dynamic underwater \gls{iout} networks employing hybrid optical--acoustic transmission. The framework consists of an \gls{eann}, which dynamically adjusts transmission power and operating frequency according to real-time channel conditions, and a \gls{dqlaecm} module that selects modulation formats and error-correction schemes based on instantaneous \gls{snr} measurements. The hybrid architecture uses optical links for short-range, high-throughput communication at distances of up to 30~m and acoustic links for long-range transmission at distances of up to 1000~m. Simulations conducted using Aqua-Sim with 100 network nodes distributed over a \(100 \times 100 \times 50\)~m volume demonstrated superior performance compared with LEACH, PEGASIS, TEEN, and MTE, achieving an energy consumption of 22~J, a \gls{ber} as low as \(10^{-8}\), a throughput of 6.2~Mbps, and a latency of 35~ms.

\gls{rl} has also been explored for routing in underwater optical wireless sensor networks. Li \textit{et al.}~\cite{li2019multiagent,li2020routing} proposed a series of \gls{marl} routing protocols for \glspl{uosn} to address dynamic topology, limited energy, and link vulnerability in underwater optical communication. Their approach models each sensor node as an independent agent that exchanges information with neighboring nodes, enabling routing decisions that account for broader network conditions rather than relying solely on local observations. The reward function considers residual energy and link quality, while an ACK-based confirmation mechanism provides negative feedback for failed transmissions. A distributed value function updates Q-values using rewards from both the selected forwarding node and the other neighboring nodes.

Their later work introduced position-based Q-value initialization and a variable learning rate to accelerate convergence. It also incorporated a global-reward mechanism that encourages nodes with higher residual energy to participate more actively in data forwarding, thereby improving network energy balance. The initial \gls{marl} protocol achieved a packet-delivery ratio of approximately 95\% and lower energy consumption than Q-learning, QDTR, and AODV. The later DMARL protocol maintained a packet-delivery ratio above 90\% and was found to be suitable for networks with fewer than 14 neighboring nodes on average.

Across these emerging applications, a notable gap is the limited number of \gls{ml} methods developed specifically for underwater optical systems. Many existing approaches are adapted from general underwater sensing, imaging, localization, or networking frameworks without explicitly accounting for optical-channel characteristics such as severe attenuation, scattering, narrow \gls{fov}, alignment sensitivity, and short communication range. This highlights the need for application frameworks designed jointly with the physical and operational characteristics of \gls{uowc}. Section~\ref{sec:challenges} discusses several recurring challenges identified across the reviewed studies.

\section{Open Challenges and Future Research Directions}
\label{sec:challenges}
\begin{figure*}[t]
    \centering
    % Source: figures/uwoc-joint-optimization.svg; regenerate the PDF with
    %   rsvg-convert -f pdf -o figures/uwoc-joint-optimization.pdf figures/uwoc-joint-optimization.svg
    \includegraphics[width=\textwidth]{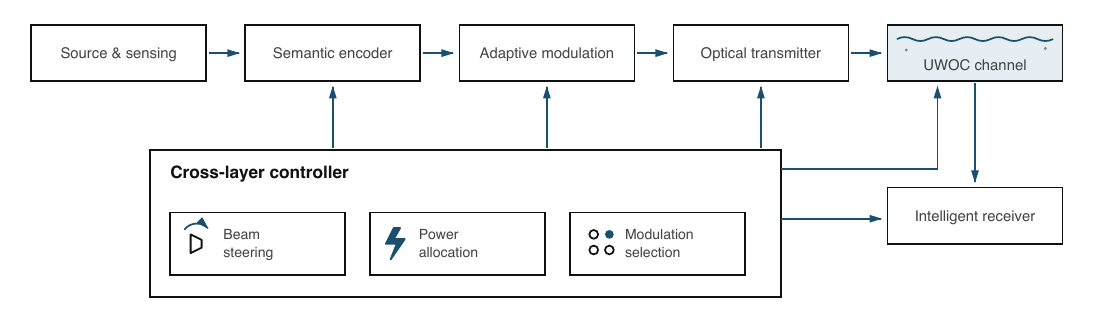}
    \caption{Conceptual architecture for joint system-level optimization in ML-enabled UOWC. The cross-layer controller coordinates semantic encoding, adaptive modulation, optical transmission, beam steering, power allocation, and receiver processing across the communication pipeline.}
    \label{fig:joint}
\end{figure*}

The preceding sections have demonstrated that machine learning has produced meaningful advances across the full \gls{uowc} pipeline. Nevertheless, the surveyed literature also reveals a consistent set of limitations that collectively constrain the deployment of \gls{uowc} systems in real underwater environments. This section identifies and discusses the principal open challenges, drawing on the specific limitations observed, and pointing to future research directions.
\label{subsec:ch_joint}
\begin{table*}[t]
\caption{Validation Modes Reported Across the Surveyed
\gls{uowc} Pipeline Stages}
\label{tab:validation_modes}
\centering
\footnotesize
\setlength{\tabcolsep}{5pt}
\renewcommand{\arraystretch}{1.15}

\begin{tabularx}{\textwidth}{
@{}>{\raggedright\arraybackslash}p{3.0cm}
*{5}{>{\centering\arraybackslash}X}@{}
}
\toprule
\textbf{UOWC Stage} &
\textbf{Simulation} &
\textbf{Experiment} &
\shortstack{\textbf{Simulation}\\\textbf{and Experiment}} &
\shortstack{\textbf{Emulated}\\\textbf{UOWC}} &
\textbf{Total} \\
\midrule

Channel
& 4 & 6 & 1 & 0 & 11 \\

Transmitter
& 9 & 2 & 1 & 3 & 15 \\

Receiver
& 6 & 7 & 1 & 0 & 14 \\

Link alignment
& 6 & 3 & 3 & 0 & 12 \\

Emerging applications
& 7 & 9 & 0 & 0 & 16 \\
\midrule

\textbf{Total}
& \textbf{32}
& \textbf{27}
& \textbf{6}
& \textbf{3}
& \textbf{68} \\
\bottomrule

\end{tabularx}
\end{table*}
\subsection{The Simulation Gap}
A major limitation across the surveyed literature is the
continued reliance on simulated training and evaluation data.
Table~\ref{tab:validation_modes} shows that simulation-only
evaluation is the largest individual validation category, accounting
for 32 of the 68 classified studies, or 47.1\%. Simulation is involved
in 38 studies, or 55.9\%, when studies combining simulation and
experiments are included. Models trained on simulated channels are expected to exhibit performance degradation when deployed on real links, where environmental conditions vary continuously~\cite{Xu2022MCimprove, Wen2023MCMPS,Geirhos_2020}.

The root cause is the distributional mismatch between simulation and reality. Real underwater optical channels involve spatially and temporally correlated turbulence, background ambient-light variability, hardware nonlinearities, and mechanical vibration from platform dynamics~\cite{11318578}.

Several directions are promising for bridging this gap. Domain adaptation and transfer learning provide principled frameworks for adapting models evaluated on simulated datasets to real deployment conditions with limited labeled data~\cite{Pan2010Transfer,8861136}. Meta-learning approaches such as \gls{maml}~\cite{pmlr-v70-finn17a}, already applied in the receiver context~\cite{9817578}, offer a particularly natural solution: by training on a distribution of simulated channel environments, the model can adapt rapidly to a new real environment with only a small number of pilot measurements.

\subsection{Real-Time Deployment }
\label{subsec:ch_realtime}

Most of the surveyed \gls{ml} frameworks are trained offline, while their computational requirements during deployment are rarely evaluated. Practical deployment on \gls{uowc} platforms, however, requires consideration of inference latency, computational complexity, memory requirements, and energy consumption. These factors become particularly important for embedded and mobile underwater platforms, where the computational resources available for real-time inference may be limited.

Several surveyed studies employ relatively complex deep-learning architectures, including the YOLOv8s--RTMPose cascade of Jia et al.~\cite{10663260}, which reports an inference time of 9.1~ms on a GPU, the Transformer-based semantic communication framework of Xu et al.~\cite{10729883}, and the multi-agent \gls{drl} beam-control framework of Shin et al.~\cite{10537612}. While model storage and FLOPs provide useful indicators of computational complexity, they do not fully characterize deployment performance. Key future directions include model compression through pruning, quantization, and knowledge distillation~\cite{cheng2020surveymodelcompressionacceleration}. Edge computing architectures may provide a practical approach for offloading computationally intensive tasks such as beam alignment and semantic encoding~\cite{9837289}.

\subsection{Generalization Across Water Types and Environments}
\label{subsec:ch_generalization}

The four canonical water types commonly considered in the \gls{uowc} literature provide useful benchmark environments~\cite{Kaushal2016,UOWC_comprehensive2025}. However, these discrete categories represent only a coarse approximation of the diversity of real underwater optical environments, where optical properties can vary continuously with environmental conditions.
The generalization of trained models across this continuous parameter space is poorly characterized in the surveyed papers. The MAML--DCGAN approach in~\cite{11027977}
demonstrated adaptation when emulating previously unseen attenuation
conditions. However, its generalization across distinct real water
bodies remains unverified. Nevertheless, underwater channel conditions can vary continuously with environmental factors such as temperature, salinity, and depth, resulting in spatially varying turbulence characteristics~\cite{Xu2023WGG}. Future work should therefore evaluate model generalization under continuous and correlated variations in optical channel parameters. Generalization frameworks such as few-shot and zero-shot learning may also be useful for this problem~\cite{pmlr-v70-oh17a,mehta2025generalizationtheoryzeroshotprediction}.
\subsection{Joint System-Level Optimization}

A recurring observation across the surveyed literature is that learning-based methods are applied to individual pipeline components in isolation: channel estimation, modulation classification, equalization, and beam alignment are each optimized separately, without coordination across stages. This modular approach is convenient but sacrifices the performance gains available from joint optimization.

The end-to-end autoencoder framework of O'Shea and Hoydis~\cite{OShea2017}, 
extended to \gls{uowc} by Zhai~\cite{10.1145/3398329.3398367} and Zou 
et al.~\cite{9514508}, represents an important step toward joint transmitter 
and receiver optimization. However, these frameworks do not jointly integrate 
other system functions such as channel estimation, beam alignment, and power 
control into the learned optimization process. Recent semantic communication 
systems~\cite{Xu:24,10729883} further integrate source and channel coding, 
with some approaches adapting semantic encoding and transmission rates to 
channel-state information~\cite{10729883}.  A fully integrated \gls{uowc} framework 
could jointly optimize source encoding, modulation, beam divergence, power 
allocation, receiver detection, and equalization, as illustrated in Fig.~\ref{fig:joint}. Differentiable channel models represent one promising approach for joint transmitter–receiver optimization with the channel in the training loop~\cite{E2Elearning2020}. The interaction between semantic communication quality and
physical-channel reliability remains largely unexplored in
\gls{uowc}.

\subsection{Physics-Informed and Hybrid Learning Approaches}
\label{subsec:ch_pinns}

The purely data-driven approaches that dominate the surveyed literature treat the underwater optical channel as a black box, ignoring its rich physical structure. This necessitates large training datasets that rediscover structures already known from physics, and the learned representations are difficult to interpret or audit~\cite{He2019ModelDriven}.

Physics-informed neural networks embed governing equations directly into the training loss, constraining the network to produce physically consistent outputs~\cite{PINNwireless2024}. Model-driven deep learning~\cite{He2019ModelDriven}, which unrolls iterative signal processing algorithms as trainable neural network layers, offers a complementary approach that has been applied to channel estimation and detection in \gls{rf} systems but not yet in \gls{uowc}.

The robustness of purely data-driven methods in dynamic and adversarial environments is also a concern~\cite{RobustDL2024}. Learning-based beam controllers trained on specific sea-state distributions may fail under unusual wave conditions; modulation classifiers trained on Gamma-Gamma fading may misclassify under \gls{egg} fading. Incorporating physical constraints into the learning formulation is one promising path toward robust, interpretable, and data-efficient \gls{uowc} systems.

\subsection{Integrated Sensing, Communication, and Power Transfer}
\label{subsec:ch_isac}

The emerging applications surveyed in Section~\ref{sec:emerging_applications} point toward a convergence of communication, sensing, localization, and power transfer in a single \gls{uowc} system. Simultaneous lightwave information and power transfer~\cite{shin2024adaptive, shang2025slipt}, integrated sensing and communication~\cite{ISAC_UWOC2024}, and joint communication and imaging~\cite{krishnan2021optical:21, joshi2024underwater} each address this convergence from a different angle.

A unified \gls{uowc} framework that simultaneously optimizes data throughput, energy harvesting efficiency, target detection probability, and node localization accuracy represents an important but highly challenging future direction. Such a joint system would require balancing multiple objectives under dynamic channel conditions and energy constraints. Building on the multi-agent \gls{drl}-based beam alignment framework of Shin et al.~\cite{10537612} and their hierarchical \gls{drl}-based \gls{slipt} framework~\cite{shin2024adaptive}, multi-agent \gls{drl} could provide a promising approach for jointly optimizing these competing objectives.

\section{Conclusion}
\label{sec:conclusion}

In this survey, we have provided a comprehensive review of the research progress of machine learning techniques in underwater optical wireless communication systems. The survey examined machine learning algorithms employed across each component of the \gls{uowc} pipeline, organized from channel modeling through to emerging applications. We first discussed machine learning methods applied to channel modeling and estimation. In the transmitter design component, we reviewed end-to-end autoencoder-based transceivers, modulation classification, link adaptation, and semantic communication systems that jointly learn source and
channel representations, with modulation handled differently across the surveyed frameworks. On the receiver side, we surveyed deep learning-based detection, demodulation, and equalization methods that compensate for hardware nonlinearities and channel impairments that resist conventional signal processing. For link alignment and beam control, we examined computer vision-based source detection methods and reinforcement learning algorithms that adapt beam divergence, orientation, and transmission power in response to dynamic underwater environments. Finally, we reviewed emerging applications of machine learning in \gls{uowc} beyond data transmission, including simultaneous lightwave information and power transfer, underwater image enhancement and object detection, and multi-sensor localization.

\bibliographystyle{IEEEtran}
\bibliography{references}

\end{document}